\documentclass[reprint,amssymb,noeprint,twocolumn,longbibliography]{revtex4-2}

\usepackage{graphicx}
\usepackage{amsfonts}
\usepackage{amsmath}

\usepackage[usenames,dvipsnames]{color}
\usepackage{bm}
\usepackage{amssymb}
\usepackage{braket}

\usepackage{multirow}
\usepackage{epstopdf}
\usepackage[normalem]{ulem}

\usepackage{hyperref}
\hypersetup{
colorlinks=true,
linkcolor=blue,
citecolor=blue,
filecolor=green,
urlcolor=blue,
}
\usepackage[nameinlink,capitalise]{cleveref}

\begin{document}

\title{Multichannel quantum defect theory with a frame transformation \\ for ultracold {atom-molecule} collisions in magnetic fields}

%  nuclear spin entanglement via long-range electric dipolar interactions of ultracold polar molecules}
%\title{Entangling nuclear spins via electric dipole-dipole interactions of ultracold polar molecules}

\author{Masato Morita$^{1}$, Paul Brumer$^{1}$, and Timur V. Tscherbul$^{2}$}

\affiliation{$^{1}$Chemical Physics Theory Group, Department of Chemistry, and Center for Quantum Information and Quantum Control, University of Toronto, Toronto, Ontario, M5S 3H6, Canada}
\affiliation{$^{2}$Department of Physics, University of Nevada, Reno, Nevada, 89557, USA}

\date{\today}

\begin{abstract}
We extend the powerful formalism of multichannel quantum defect theory combined with a frame transformation (MQDT-FT) to ultracold 
{atom-molecule} %molecular 
collisions in magnetic fields. By solving the coupled-channel equations with hyperfine and Zeeman interactions omitted at short range, MQDT-FT enables a drastically simplified description of the intricate quantum dynamics of ultracold molecular collisions in terms of a small number of short-range parameters. We apply the formalism to ultracold Mg + NH collisions in a magnetic field, achieving a 10$^4$-fold reduction in computational effort.
\end{abstract}
\maketitle

\newpage

Recent advances in cooling and trapping diatomic and polyatomic molecules have established ultracold molecular gases as an emerging platform for quantum information science, ultracold chemistry, and precision searches for new physics beyond Standard Model \cite{Carr_09,Balakrishnan:16,Bohn_17,Demille_17,Devolder:20,Devolder:21}.
The exquisite control over molecular degrees of freedom achieved in these experiments enables the exploration of novel regimes of ultracold chemical dynamics  tunable by external electromagnetic fields \cite{Lemeshko_13,Krems_09,Dulieu_17}. 
However, the use of molecules for these applications has been hindered by rapid losses observed in ultracold molecular gases due to the formation of intermediate complexes in two-body collisions \cite{Mayle_12,Mayle_13,Christianen_19,Gregory_19,Liu_20,Gersema_21,Nichols_22,Bause_23}.
Therefore, understanding and controlling ultracold molecular collisions have long been recognized as major goals in the field \cite{Carr_09,Balakrishnan:16,Bohn_17}.
This, however, has proven to be an elusive task due to the enormous number of molecular states (rotational, vibrational, fine, and hyperfine) strongly coupled by highly anisotropic interactions and external fields, which results in skyrocketing computational costs of rigorous coupled channel (CC) calculations on ultracold molecular collisions \cite{Morita_19,Morita_20}. 
%The rigorous coupled channel (CC) method for solving the time-independent Schr\"{o}dinger equation is a powerful tool to study ultracold molecular collisions \cite{Morita_19,Morita_20}. However, CC calculations struggle to obtain converged results owing to the need of including all of these states, which results in skyrocketing computational costs.
While efficient basis sets have been developed to mitigate this problem \cite{Tscherbul:10,Suleimanov:12,Morita_18,Morita_20,Tscherbul_23}, fully converged CC calculations including all degrees of freedom are yet to be reported on ultracold K~+~NaK \cite{Yang_19,Wang_21},  Na~+~NaLi \cite{Son_22,Park:23b}, and Rb~+~KRb \cite{Nichols_22} collisions explored in recent experiments.

A promising avenue toward resolving these difficulties could be based on multichannel quantum defect theory (MQDT), an elegant technique for solving the time-independent Schr\"{o}dinger equation based on the separation of distance and energy scales in ultracold collisions \cite{Greene_82,Mies_84,Gao_96,Gao:09,Gao:11,Mies_00,Raoult_04,Croft_11,Croft_12,Croft_13,Jisha_14,Jisha_14b}.
MQDT allows one to avoid the costly numerical procedure of solving CC equations over extended ranges of the radial coordinate $R$, collision energy, and magnetic field, leading to a substantial reduction of computational cost \cite{Croft_11,Croft_12, Croft_13,Jisha_14,Jisha_14b,Burke_98,Burke_99,Gao_05}.
However, in conventional MQDT, it is still necessary to solve CC equations in the short-range collision complex region, which is forbiddingly difficult for the atom-molecule systems interacting via deep and strongly anisotropic potentials, such as those studied in recent experiments \cite{Yang_19,Wang_21,Son_22,Park:23b,Nichols_22}.

MQDT becomes especially powerful when combined with a frame transformation (FT) approach, in which the hyperfine-Zeeman structure of colliding atoms is neglected at short range.  This results in a large reduction of computational effort and provides a physically meaningful description of ultracold atomic collisions in terms of a few short-range parameters \cite{Burke_98,Burke_99,Gao_05, Hanna_09,Idziaszek_2011,Perez-Rios_15,Li_15,Giannakeas_13}. An extension of MQDT-FT to the much more computationally intensive (and less well understood) ultracold molecular collisions would thus be highly desirable. Previous theoretical work has shown that  MQDT can be successfully applied to describe ultracold atom-molecule collisions \cite{Croft_11,Croft_12,Croft_13,Jisha_14} and chemical reactions \cite{Jisha_14b}, but these calculations did not consider the essential FT aspect of MQDT, which makes it such an indispensable tool in modern ultracold atomic collision theory  \cite{Burke_98,Burke_99,Gao_05, Hanna_09,Idziaszek_2011,Perez-Rios_15,Li_15}.
The application of MQDT-FT to atom-molecule collisions faces significant challenges due to uniquely complex features of molecular structure such as rotational states \cite{Brown:03,DeMille:15}, multiple nuclear spins, and inherently anisotropic atom-molecule interactions \cite{Carr_09,Balakrishnan:16,Bohn_17}.
% and  interact with intramolecular spin degrees of freedom.
% in ways foreign to atoms.
%Furthermore, determining a simplified yet effective Hamiltonian to describe the short-range dynamics in atom-molecule collisions is neither straightforward nor unique. Consequently, it becomes imperative to devise a tailored MQDT-FT formulation and assess its validity through numerical calculations.
\color{black}

Here, we show that the MQDT-FT approach can be extended to ultracold molecular collisions in a magnetic field, enabling one to drastically simplify their rigorous theoretical description. This is achieved by using compact CC basis sets at short range, which exclude the hyperfine and Zeeman interactions. These interactions are incorporated at long range via MQDT-FT boundary conditions, resulting in a complete description of ultracold atom-molecule collision dynamics across large collision energy and magnetic field ranges in terms of a small number of short-range parameters. 
These parameters can be used to, e.g., fit experimental observations and to obtain insight into complex molecular collision dynamics without performing expensive CC calculations. 
Our results show the potential of MQDT-FT to significantly extend the scope of ultracold atom-molecule collisions and chemical reactions amenable to rigorous quantum dynamical studies, which would facilitate the accurate characterization of atom-molecule Feshbach resonances observed in recent pioneering experiments \cite{Yang_19,Wang_21,Son_22,Park:23b}, as well as new insights into quantum chaotic behavior and microscopic interactions in atom-molecule collision complexes \cite{Croft_17}. 

{\it Theory.} We begin by briefly reviewing the key concepts of MQDT as they apply to ultracold atom-molecule collisions in a magnetic field.
%\color{red}
%MQDT-FT for atom-molecule collisions.
The Hamiltonian of the atom-molecule collision complex 
%\color{red} 
%in a magnetic field 
\color{black}
is (in atomic units) \cite{Krems_04}
\begin{equation}
{
\hat{H} = - \frac{1}{2\mu R} \frac{\partial^2}{\partial R^2}R + \frac{\hat{\mathbf{l}}^2}{2\mu R^2} +\hat{H}_\text{as} + \hat{V}(R,\theta)
}.
\label{eq:Heff}
\end{equation}
Below we will use ultracold Mg($^1$S)~+~NH($\tilde{X}^3\Sigma^-$) collisions as a representative example, which served as a testbed for applying MQDT to ultracold atom-molecule collisions \cite{Croft_11}. 
%We note, however, that our treatment can be readily generalized to collisions of atoms and molecules with a more complex internal structure, and to molecule-molecule collisions. 
In \cref{eq:Heff}, $\mu$ and $\hat{\mathbf{l}}$ are the reduced mass and the orbital angular momentum for the collision, $R$ is the atom-molecule distance, and  $\hat{H}_\mathrm{as}$ is the asymptotic Hamiltonian
%($\mu=9.60046$ amu), 
\begin{equation}
\hat{H}_\text{as} = 
B_{e}\,\hat{\mathbf{N}}^2+ \hat{H}_\text{fs} + \hat{H}_\text{hfs}  + \hat{H}_\text{Z},
\label{eq:Has}
\end{equation}
%where $B_\mathrm{rot}$ is the rotational constant ($B_\mathrm{rot}=16.2712$ cm$^{-1}$), $\gamma_\mathrm{sr}$ is the spin-rotation constant ( $\gamma_\mathrm{sr}=-0.0546$ cm$^{-1}$), and $\lambda_{ss}$ is the spin-spin constant ($\lambda_{ss}=0.9199$ cm$^{-1}$) \cite{Bizzocchi_2018}. 
where $B_{e}$ is the rotational constant,  $\hat{\mathbf{N}}$ is the rotational angular momentum of the molecule, $\hat{H}_\text{fs} = \gamma_\mathrm{sr}\,\hat{\mathbf{N}}\cdot\hat{\mathbf{S}}+\frac{2}{3}\lambda_\mathrm{ss}\sqrt{\frac{24\pi}{5}}\sum^{}_{q}(-1)^q\,Y_{2\,-q}(\hat{\bm{r}})[\hat{\mathbf{S}} \otimes \hat{\mathbf{S}}]_q^{(2)}$  is the intramolecular fine-structure Hamiltonian, which depends on the electron spin $\hat{\mathbf{S}}$ 
%($S=|\hat{\mathbf{S}}|=1$ for $^3\Sigma$ molecules) 
and the orientation of the molecular axis $\bm{r}$, and $\gamma_\text{sr}$ and $\lambda_\text{ss}$ are the spin-rotation and spin-spin interaction constants \cite{Wallis_09,Maykel_11}.  
% are the intramolecular spin-rotation and spin-spin constants, respectively. 
%For these constants, we employ the same values used in previous studies \cite{Maykel_11,Croft_11} (see also supplemental material \cite{SM}). 
The Zeeman interaction $\hat{H}_\text{Z} = g_S \mu_\mathrm{B} \mathbf{B}\cdot \hat{\mathbf{S}}$, where $g_S \simeq 2.002$ is the electron $g$-factor, $\mu_\mathrm{B}$ is the Bohr magneton, the magnetic field vector $\mathbf{B}$ defines the space-fixed quantization axis $z$, and the hyperfine interaction $\hat{H}_\text{hfs}$ is considered below. 
We use the %accurate 
{\it ab initio} interaction potential $V(R,\theta)$ for Mg+NH \cite{MOLSCAT}, which depends on $R$ and the %atom-molecule 
Jacobi angle $\theta$  \cite{SM}.
% (see the Supplemental Material for computational details \cite{SM}).

In MQDT, the matrix solution $\bm{\Psi}$ of the Schr\"odinger equation $\hat{H}\bm{\Psi}=E\bm{\Psi}$ in the basis of eigenvectors of the asymptotic Hamiltonian [\cref{eq:Has}] is matched to \cite{Mies_00,Croft_11} 
\begin{equation}
\label{eq:MQDT}
\bm{\Psi}( R_{m} )= R_{m}^{-1}\,[\bm{f}(R_{m})+\bm{g}(R_{m})\,\bm{K}^{\text{sr}}\,],
\end{equation}
where the matching radius $R=R_m$ marks the boundary between the short-range and long-range regions, $E$ is the total energy, $\bm{K}^{\text{sr}}$ is the short-range $K$-matrix, which includes open and weakly closed channels, and $\bm{f}(R)$ and $\bm{g}(R)$ are  diagonal matrices of regular and irregular reference solutions \cite{SM}.
The transition $T$-matrix and state-to-state integral cross sections are obtained from $\bm{K}^{\text{sr}}$ following the standard sequence of MQDT steps, which involves taking into account the coupling between the open and weakly closed channels, and the threshold effects \cite{SM}. 
Because the $N_\text{ref}$ reference channels are decoupled at $R\ge R_m$, this procedure is computationally efficient, scaling  linearly with  $N_\text{ref}$.
%number of reference channels.
\color{black}

In the spirit of MQDT-FT for atomic collisions  \cite{Burke_98,Burke_99,Gao_05, Hanna_09,Idziaszek_2011,Perez-Rios_15,Li_15}, we seek to describe short-range quantum dynamics with a simplified Hamiltonian $\hat{H}_\mathrm{0}$ obtained by omitting the terms, which are small compared to $\hat{V}$, from $\hat{H}_\text{as}$ in \cref{eq:Has}. 
We then solve the Schr\"odinger equation $\hat{H}_\text{0}\bm{\Psi}_0=E\bm{\Psi}_0$ subject to the boundary conditions
\cite{Mies_00,Croft_11} 
\begin{equation}
\label{eq:MQDT-FT}
\bm{\Psi}_0( R_{m} )= R_{m}^{-1}\,[\bm{f}(R_{m})+\bm{g}(R_{m})\,\bm{K}_\text{0}^{\text{sr}}\,]
\end{equation}
to obtain  $\bm{K}_\mathrm{0}^\text{sr}$, the  MQDT-FT analog of the short-range $K$-matrix.
Note that $\hat{H}_0$ only occurs in MQDT-FT, but not in conventional MQDT.
%The occurrence of  marks the primary difference between conventional MQDT (in which short-range physics is treated exactly) and   and MQDT-FT, 
% The choice of $\hat{H}_\text{0}$ thus plays a key role in determining the accuracy and efficiency of MQDT-FT.
 In atomic MQDT-FT,  $\hat{H}_\text{0}$ is typically chosen by omitting the hyperfine and Zeeman interactions of the collision partners from $\hat{H}$ \cite{Burke_98,Burke_99,Gao_05, Hanna_09,Idziaszek_2011,Perez-Rios_15,Li_15}. By contrast, as shown below, in atom-molecule MQDT-FT it is possible to choose $\hat{H}_0$ in several nonequivalent ways  depending on the nature of the collisional transition of interest, demonstrating the flexibility of the technique. 
The exact short-range $K$-matrix ($\bm{K}^\text{sr}$) defined by \cref{eq:MQDT} is then  approximated as
% applying a FT to $\bm{K}_\mathrm{0}^{\text{sr}}$,
\color{black}
\begin{equation}\label{eq:FT}
\bm{K}^\text{sr} \simeq \bm{U}^{\dagger}\bm{K}_\mathrm{0}^{\text{sr}}\bm{U},
\end{equation}
where $\bm{U}$ is 
an orthogonal matrix composed of the eigenvectors of the asymptotic Hamiltonian (\cref{eq:Has})  \cite{SM}. 
%$\hat{H}_\text{as}$ of $\hat{H}$.  % in
%As discussed below and in the Supplemental Material \cite{SM}, $\bm{K}_\mathrm{0}^{\text{sr}}$ is firstly obtained in the basis of the eigenvectors of the asymptotic Hamiltonian $\hat{H}_\text{0,as}$ of $H_\text{0}$ ($H_\text{0} \to H_\text{0,as}$ as $R\to\infty$).
\color{black}
Note that $\bm{K}_\mathrm{0}^{\text{sr}}$ is much easier to compute than $\bm{K}^\text{sr}$ 
because $\hat{H}_0$ can be chosen in such a way as to exclude the small hyperfine and Zeeman interactions, %(see below)
thus eliminating the need to use enormous CC basis sets including the electron and/or nuclear spin basis functions of the collision partners \cite{Tscherbul_23}. In addition, since $\hat{H}_0$ is independent of external fields, $\bm{K}_\mathrm{0}^{\text{sr}}$ can be efficiently computed using the total angular momentum (TAM) basis as shown below, and it is independent of $M$, the $z$ projection of $J$.

{ {\it Results.} } 
We now apply MQDT-FT to describe the quantum dynamics of ultracold  Mg~+~NH collisions in a magnetic field, focusing on the inelastic transitions between the different Zeeman sublevels of NH($N=0$) labeled by the $z$-projection of the molecule's electron spin $M_S=-1,0,1$. These transitions cause trap loss and thus limit the efficiency of sympathetic cooling of NH molecules by Mg atoms \cite{Wallis_09}. 
For now, we neglect the hyperfine interaction ($\hat{H}_\text{as} = B_e\hat{\bm{\mathrm{N}}}^2 + \hat{H}_\text{fs}+ \hat{H}_\text{Z}$) and choose the simplified short-range Hamiltonian $\hat{H_0}$ 
%to compute $\bm{K}_\mathrm{0}^{\text{sr}}$
by neglecting the Zeeman interaction as  
%$\hat{\mathcal{H}}_{0} = \hat{\mathcal{H}} - \hat{\mathcal{H}}_\text{Z}$.
$\hat{H}_{0} = \hat{H} -\hat{H}_\text{Z}$.
Since $\hat{H}_{0}$ is field-independent, we obtain $\bm{K}_\mathrm{0}^{\text{sr}}$ by solving CC equations in the TAM 
basis  $|(NS)jlJM\rangle$, where  $\hat{\mathbf{J}}=\hat{\mathbf{l}} + \hat{\mathbf{j}}$  is the total angular momentum of the atom-molecule system and $\hat{\mathbf{j}}=\hat{\mathbf{N}} + \hat{\mathbf{S}}$ is that of the diatomic molecule.
Because $\hat{H}_{0}$ is block diagonal in $J$ and independent of $M$, 
%their number is much smaller (XXX) than that used in previous MQDT calculations on Mg~+~NH, providing an $\simeq XXX$ reduction in computational cost.
the computational cost of MQDT-FT is much smaller than that of  conventional MQDT \cite{Croft_11}. 

%On the other hand, the CC equation for MQDT is composed of 534 channels for $M=1$ with the same basis set. We note that no additional short-range CC calculations are required for MQDT-FT even for the calculations with different $M$ values as long as $J_\text{max}$ is sufficiently large. 

%%%%%%%%%%%%%%%%%%%%%%%%%%%%%%%%%%%%%%%%%%%%%%%%%%%%%%
%\begin{figure}[bh!]
\begin{figure}[t]
\begin{center}
\includegraphics[height=0.162\textheight, trim = 10 0 0 0]{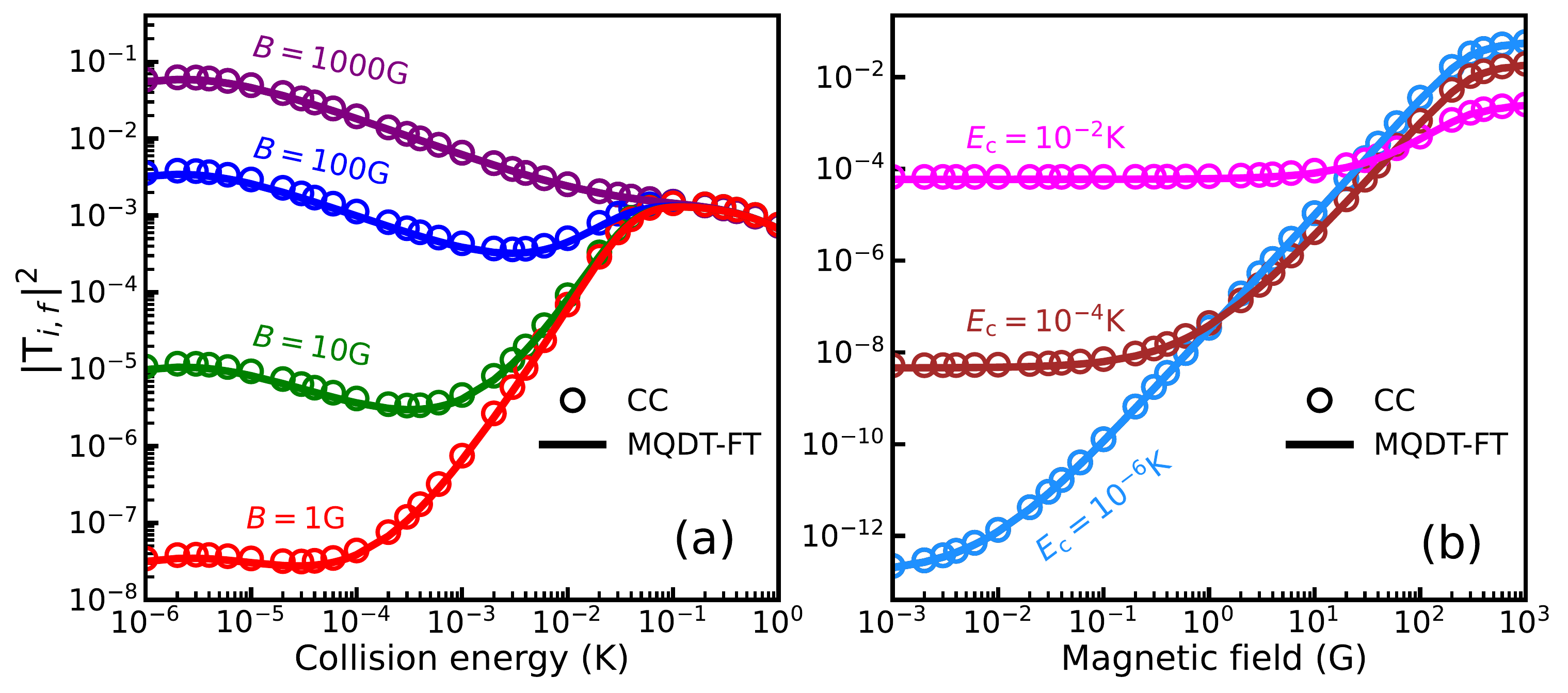}
\end{center}
\vspace{-0.5cm}
\caption{
Probabilities for the spin relaxation transition $M_S=1$ ($l=0$) $\to$ $M^\prime_S=0$ ($l^\prime=2$) in cold  Mg~+$^{14}$NH($N=0$) collisions plotted as a function of collision energy (a) and magnetic field (b). 
Solid lines -- MQDT-FT calculations, open circles -- exact CC results. }
\label{Fig_1}
\end{figure}
%%%%%%%%%%%%%%%%%%%%%%%%%%%%%%%%%%%%%%%%%%%%%%%%%%%%%%

%square of the modulus of the $T$-matrix element, $|T|^2$, for the
%partial wave-resolved 
\Cref{Fig_1} compares the probabilities $|T_{i\to f}|^2$ for the spin relaxation transition $M_S=1 \to 0$ between the Zeeman levels of NH ($N=0$) in cold Mg+NH collisions calculated using MQDT-FT with exact CC results.
Encouragingly, MQDT-FT provides an essentially exact description of the transition probability as a function of collision energy {\it and} external magnetic field with only 139 channels as compared to 954 channels required in standard MQDT \cite{Croft_12},  
%\color{magenta}
%{\bf [Sentence below is valid if "(case 2)" is selected as a reply to the Comment A4 in the reply letter]}
a $\simeq$7-fold reduction in the number of channels. Because the computational cost of solving CC equations scales as $N^3$ with the number of channels $N$, MQDT-FT offers a $\simeq$320-fold increase in computational efficiency over standard MQDT. 
%This choice of $\hat{H}_0$ is particularly advantageous in the presence of hyperfine structure, where it leads to even larger computational savings (see below).
\color{black}

\begin{figure}[t!]
\begin{center}
\includegraphics[height=0.25\textheight,keepaspectratio]{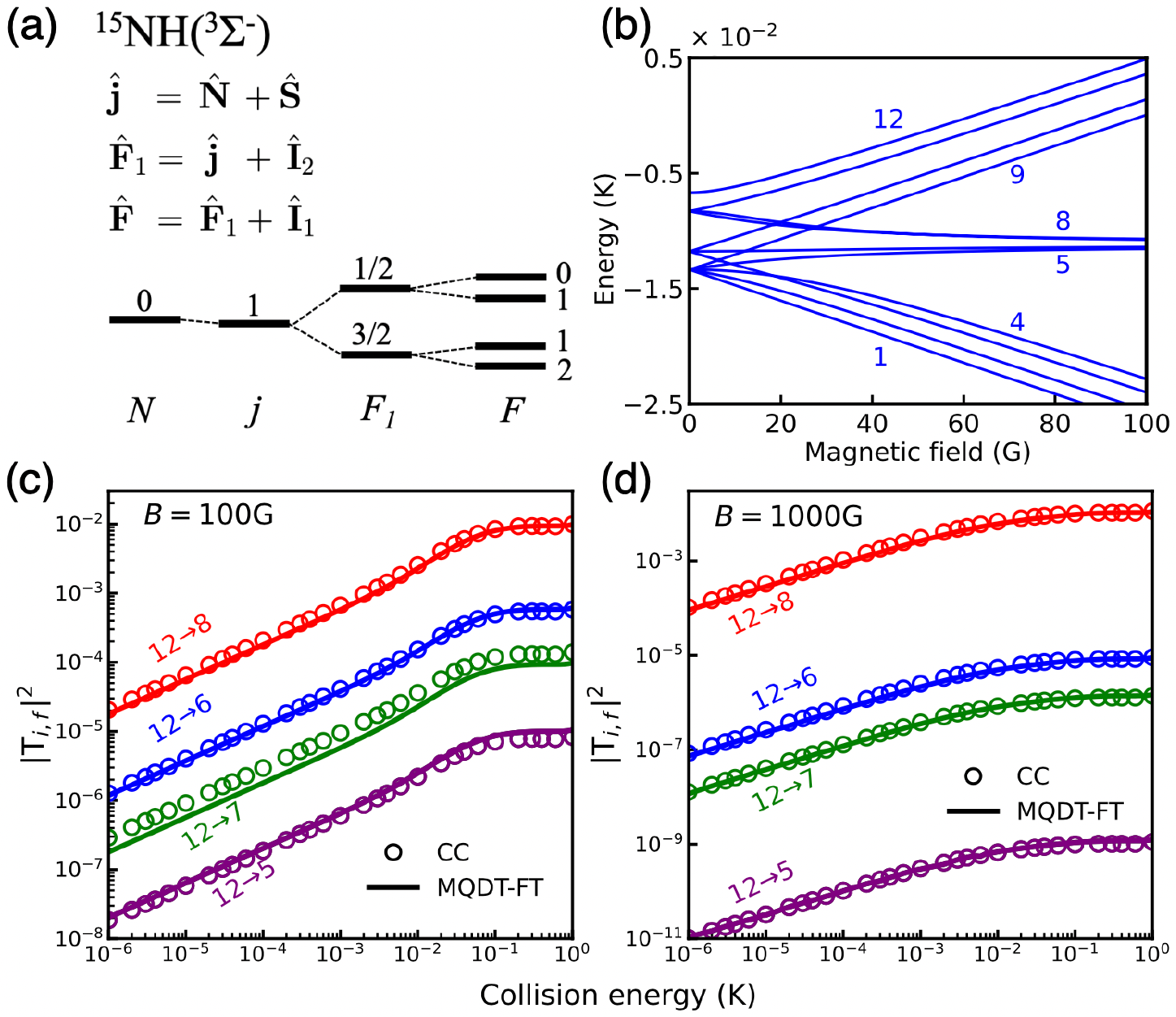}
\end{center}
\caption{Hyperfine energy levels of $^{15}$NH ($N=0$) at zero field (a) and as a function of magnetic field (b). 
%Some numbers displayed in the panel show the indexes of the energy levels in the ascending order of their energies at each magnetic field.
State-to-state transition probabilities $|T_{il, fl'}|^2$  plotted as a function of collision energy at $B=100$\,G (c) and $B=1000$\,G (d) for  Mg~+~NH($N=0$) collisions for $l=0 $ and $l^\prime=2$. 
Solid lines -- MQDT-FT calculations, %symbols 
open circles 
\color{black}
-- exact CC results. 
The initial and final states of NH are indicated next to each curve as $i \to f$ [see panel (b)].
% Transitions from the 12th state (the highest energy state in the $N=0$ manifold) are shown as $12 \to 5$ (purple), $12 \to 6$ (blue), $12 \to 7$ (green), and $12 \to 8$ (red). 
}
\label{Fig_2}
\end{figure}

The excellent performance of MQDT-FT illustrated in \cref{Fig_1} validates the assumption that $\hat{H}_\text{Z}$ plays a negligible role in short-range dynamics. 
%The effect of $\hat{\mathcal{H}}_\text{Z}$ is included indirectly via the long-range part.
The strong magnetic field dependence of the inelastic probability at 1~$\mu$K  thus arises entirely from long-range physics. Specifically, inelastic scattering is suppressed by the centrifugal barrier in the outgoing $d$-wave channel at low magnetic fields \cite{Volpi_02,Campbell_09}. 
This is an example of physically meaningful insight into ultracold atom-molecule collisions gained from MQDT-FT simulations. 
We next address the question of whether MQDT-FT can be used to describe hyperfine interactions, which play a crucial role in ultracold molecular collisions \cite{Balakrishnan:16,Bause_23}, yet are notoriously difficult to describe using standard CC methods, requiring enormous basis sets \cite{Tscherbul_23}. 
As depicted in ~\cref{Fig_2}(a) the hyperfine structure of $^{15}$NH arises from the nuclear spins of $^{15}$N ($I_1=1/2$) and H ($I_2=1/2$) which couple to $\hat{\mathbf{j}}$ 
(see above) to produce the total angular momentum $\hat{\mathbf{F}}$ of NH, which is a good quantum number at low $B$-field. The hyperfine structure is described by the Hamiltonian 
%  In this case, the nuclear spin quantum numbers for N ($I_\text{N}$) and H ($I_\text{H}$) in the NH molecule are both 1/2, and the Hamiltonian for the collision complex is given as the sum of $\hat{\mathcal{H}}$ in \cref{eq:Heff} and the hyperfine interactions $\hat{\mathcal{H}}_\text{hf}$ given as \cite{Tscherbul_07}
%\begin{equation}
%\hat{\mathcal{H}}_\text{hf} & = \sum_{i} \{ (b_i+\frac{c_i}{3}) \hat{\bm{I}_i} \cdot \hat{\bm{s}} \\
%&+\frac{c_i\sqrt{6}}{3}\sqrt{\frac{4\pi}{5}} \sum_{q=-2}^{2} (-)^q Y_{2\, -q}(\theta_r,\phi_r)[\hat{\bm{I_i}} \otimes \hat{\bm{s}} ]_q^{(2)} \},
$\hat{H}_\text{hf} = \sum_{i=1,2}  a_i \hat{\mathbf{I}}_i \cdot \hat{\mathbf{S}} + \hat{H}_\text{ahf},$ 
%\label{eq:hf}
%\end{equation}
where 
$\hat{H}_\text{ahf}=\sum_i \frac{c_i\sqrt{6}}{3}\sqrt{\frac{4\pi}{5}} \sum_{q} (-1)^q Y_{2\, -q}(\bm{r})  [\hat{\mathbf{I}}_i \otimes \hat{\mathbf{S}} ]_q^{(2)}$ is the anisotropic hyperfine interaction, and 
\color{black}
$a_i$ and $c_i$ are the isotropic and anisotropic hyperfine  constants for the $i$-th nucleus  \cite{Bailleux_12,Bizzocchi_18,SM}.  %
%, and $\hat{H}_\text{ahf}=\sum_i \frac{c_i\sqrt{6}}{3}\sqrt{\frac{4\pi}{5}} \sum_{q=-2}^{2} (-1)^q Y_{2\, -q}(\bm{r})  [\hat{\mathbf{I}}_i \otimes \hat{\mathbf{S}} ]_q^{(2)}$ is the anisotropic hyperfine interaction.
%the Fermi contact and anisotropic (tensor) hyperfine constant
%\textcolor{red}{
%Frosch and Foley parameters \cite{Frosch_52} for the $i$-th nucleus, and $b_i+{c_i/3}$ is the constant for the isotropic Fermi contact interaction and 2nd term is anisotropic tensor 
%}
%where $i=1$ and $2$ for N and H, respectively.
 %We use the constants, including the hyperfine constants $b_i$ and $c_i$ from ~\cite{Bizzocchi_18} (see also \cite{SM}).  
In the high $B$-field limit, the 12 hyperfine levels of $^{15}$NH are arranged in three groups, with 4 states per group, according to the value of $M_S = -1,0,$ and $1$, as shown in \cref{Fig_2}(b).

%%%%%%%%%%%%%%%%%%%%%%%%%%%%%%%%%%%%%%%%%%%%%%%%%%%%%%
%\begin{figure}[b!]
%The results calculated with the coupled-channel (CC) calculation and the MQDT-FT method are shown 
%%%%%%%%%%%%%%%%%%%%%%%%%%%%%%%%%%%%%%%%%%%%%%%%%%%%%%

To account for the hyperfine structure, we perform MQDT-FT calculations using the same simplified short-range Hamiltonian $\hat{H}_\mathrm{0}$ and the TAM basis set as described above without extra computational costs.
It is only necessary to augment $\bm{K}_\mathrm{0}^\text{sr}$ by the nuclear spin basis states $\ket{I_1M_{I_1}} \ket{I_2 M_{I_2}}$ before applying
the FT \cite{SM}.
%which involve the nuclear spin basis functions
% $\bm{I}$ is a 4x4 unit matrix corresponding to the space composed of 4 possible linear independent nucleus spin states such as $M_{I_\text{N}} =M_{I_\text{H}} =1/2$, $M_{I_\text{N}} =1/2$ and $M_{I_\text{H}} =-1/2$, $M_{I_\text{N}}=-1/2$ and $M_{I_\text{H}} =1/2$, and $M_{I_\text{N}}=M_{I_\text{H}} =-1/2$. 

%and the basis set for the short-range CC calculations are the same as the above calculations for \cref{Fig_1}, thus we can start from the same $\bm{K}_\mathrm{0}^\text{SR}$ without any additional calculations.
% On the other hand, $\bm{K}_\mathrm{0}^\text{SR}$ does not have a dimension for the nuclear spin degrees of freedom, thus we need to redefine $\bm{K}_\mathrm{0}^\text{SR}$ by extending as $\bm{K}_\mathrm{0}^\text{SR}\otimes\bm{I}$, where $\bm{I}$ is a 4x4 unit matrix corresponding to the space composed of 4 possible linear independent nucleus spin states such as $M_{I_\text{N}} =M_{I_\text{H}} =1/2$, $M_{I_\text{N}} =1/2$ and $M_{I_\text{H}} =-1/2$, $M_{I_\text{N}}=-1/2$ and $M_{I_\text{H}} =1/2$, and $M_{I_\text{N}}=M_{I_\text{H}} =-1/2$. 

\Cref{Fig_2}(c) and (d) show transition probabilities between the different hyperfine levels of NH in ultracold collisions with Mg atoms. 
%as functions of collision energy for the partial wave resolved ($l=0 \to l^\prime=2$) state-to-state hyperfine relaxations from the energetically highest hyperfine state (12th state) to the final states (5-8th states) in the rotational ground state ($N=0$) with the magnetic field of $B=100$G and $B=1000$G. 
We observe very good agreement between MQDT-FT and exact CC results, which indicates that the effects of hyperfine structure on ultracold atom-molecule collisions can be accurately described by MQDT-FT. 
Importantly, only 147 channels %with $J_\text{max}=7$ 
need to be coupled in our MQDT-FT calculations, whereas full CC computations involve as many as 3854 channels \cite{Maykel_11}.
Thus, MQDT-FT affords a massive reduction of computational cost by the factor of {$(3854/147)^3\simeq 1.8\times 10^4$.}
%with the converged fully uncoupled basis set  
%Given the $O(N^3)$ scaling of the computational effort with the number of channels $N$ 
%\color{red}
%Given the cubic scaling of the computational effort with the number of channels in solving CC equation,
%\color{black}

%\textcolor{red}{
%The required computational time for the short-range CC calculations can be more than 10,000 times faster than previous MQDT studies.
%}
% The computational cost of short-range This further efficiently yet precisely without using the information of $\hat{\mathcal{H}}_\text{Z}+\hat{\mathcal{H}}_\text{hf}$ in the short-range. 
%While the above calculations are efficient enough to address unexplored relevant molecular collisions, we explicitly treat multiple open channels simultaneously. 

An attractive feature of the MQDT-FT approach is its capability to describe the complex quantum dynamics of ultracold collisions in terms of a small number of ''universal'' parameters \cite{Burke_99,Gao_05}.  While this capability has provided invaluable physical insight into ultracold atomic collisions  \cite{Burke_99,Gao_05}, it has been unclear whether it can be extended to more complex molecular collisions. 
A minimal MQDT-FT model of ultracold Mg~+~NH collisions is based on a $2\times 2$ short-range $K$-matrix obtained 
using only the elements of $\bm{K}^\text{sr}$ corresponding to the initial and final channels ($K_{ii}^\text{sr}$, $K_{if}^\text{sr}=K_{fi}^\text{sr}$, and $K_{ff}^\text{sr}$).
%only the four elements associated with the initial $i$ and final $f$ channels in $\bm{K}^\text{SR}$, namely $K_{ii}^\text{SR}$, $K_{if}^\text{SR}=K_{fi}^\text{SR}$, and $K_{ff}^\text{SR}$, which was computed in the calculation for \cref{Fig_1}.

\begin{figure}[t]
\begin{center}
\includegraphics[height=0.17 \textheight,keepaspectratio]{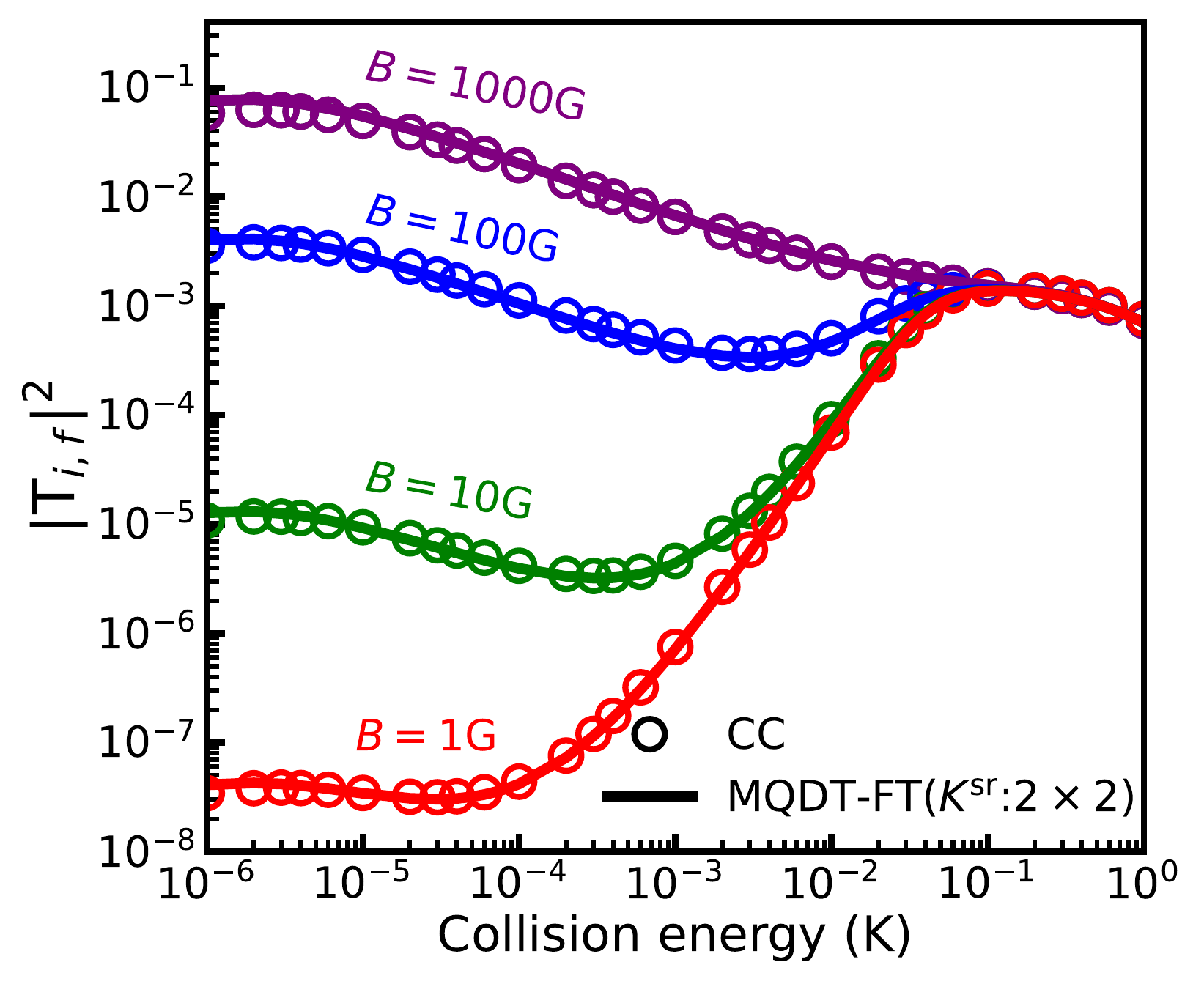}
\end{center}
\vspace{-0.15cm}
\caption{Probabilities for the transition $M_S=1$ ($l=0$) $\to$ 
$M^\prime_S=0$ ($l^\prime=2$) in cold Mg~+~$^{14}$NH$(N=0)$ collisions as a function of collision energy. Solid lines -- MQDT-FT calculations based on a $2\times2$ short-range $K$-matrix ($K^\text{sr}$:$2\times2$), open circles -- exact CC results. 
}
\label{Fig_3}
\end{figure}

As shown in \cref{Fig_3}, the probability for the $M_S=1$ $\to$ $M^\prime_S=0$ transition obtained with the two-channel model is in excellent agreement with the full 954-channel results. {\it Remarkably, the quantum dynamics of inelastic spin relaxation in ultracold Mg~+~NH collisions in a magnetic field can be described nearly exactly using just three short-range parameters ($K_{ii}^\text{sr}$, $K_{if}^\text{sr}$, and $K_{ff}^\text{sr}$) over a wide range of collision energies.} 
This result highlights the universal nature of complex ultracold atom-molecule collisions in a dc magnetic field, and suggests MQDT-FT as a possible alternative to the approaches based on a single-channel universal model (UM) \cite{Idziaszek_10,Frye_15}.
While the phenomenological $y$ parameter of the UM can be related to the results of CC calculations only in some special cases \cite{Frye_15}, the MQDT-FT parameters 
(${K}^\text{sr}_{ij}$)
are readily available from modest CC calculations, which only need to be performed in the short-range region, and do not need to include the Zeeman and hyperfine interactions.

%%%%%%%%%%%%%%%%%%%%%%%%%%%%%%%%%%%%%%%%%%%%%%%%%%%%%%
%\begin{figure}[b!]
%%%%%%%%%%%%%%%%%%%%%%%%%%%%%%%%%%%%%%%%%%%%%%%%%%%%%%

{We next} consider the possibility of MQDT-FT based on an even more drastically simplified short-range Hamiltonian $\hat{H}_0$,  {\it which includes only rovibrational degrees of freedom}. 
%We now consider the possibility of more efficient MQDT-FT
%based on an even more drastically simplified short-range Hamiltonian $\hat{H}_0$,  {\it which includes only rovibrational degrees of freedom}.
{The flexibility in choosing $\hat{H}_0$ is an essential new aspect of atom-molecule MQDT as opposed to atomic MQDT-FT \cite{Burke_98,Burke_99,Gao_05, Hanna_09,Idziaszek_2011,Perez-Rios_15,Li_15}.}
Because of the absence of all spin-dependent interactions in $\hat{H}_0$, it cannot describe inelastic transitions between the different hyperfine-Zeeman sublevels of NH($N=0$), which, for collisions with structureless atoms such as Mg, are mediated by intramolecular spin-dependent interactions \cite{Krems_04,Campbell_09}. 
We thus focus on inelastic transitions between the different fine-structure components of excited rotational states of NH.
We obtain $\bm{K}^\mathrm{sr}_0$ %by solving CC equations 
%for the simplified short-range Hamiltonian $\hat{H}_0$  ($\hat{H}_0  \to  B_e\hat{\bm{\mathrm{N}}}^2$ in the limit $R \to \infty$)
%is   $\hat{H}_0 = \hat{H} - \hat{H}_\text{fs} - \hat{H}_\text{hfs}-\hat{H}_\text{Z}$ with 
 by solving CC equations in the basis set $|(Nl)J_rM_{r}\rangle$, where $\hat{\mathbf{J}}_r=\hat{\mathbf{N}} + \hat{\mathbf{l}}$  is the total rotational angular momentum (TRAM) of the collision complex \cite{Tscherbul_23}.
The basis functions $|(Nl)J_rM_{r}\rangle$ are the eigenstates of $\hat{H}_\mathrm{0,as}$, the asymptotic Hamiltonian part of $\hat{H}_\mathrm{0}$. 
%the procedures described in Ref.~\cite{Corey_83}. Here, $\hat{\mathcal{H}}$ is equal to the right-hand-side in \cref{eq:Has}, and $\hat{\mathcal{H}}_\mathrm{0}$
%as described above
%and solving solving the CC equations for the resulting spinless atom-molecule Hamiltonian  obtained from  defined by neglecting the spin degrees of freedom from $\hat{\mathcal{H}}$ (see \cref{eq:Hmol}) as
%\begin{equation}
%\hat{\mathcal{H}}_\mathrm{0} = - \frac{1}{2\mu R} \frac{d^2}{dR^2}R + \frac{\hat{\bm{l}}^2}{2\mu R^2} +B_\mathrm{rot}\,\hat{\bm{N}}^2+ V_\text{int}(R,\theta)
%\label{eq:Hsf}
%\end{equation}
%using the basis set $|J_rM_{J_r}(Nl)\rangle$, where $\bm{J}_r=\bm{N}+\bm{l}$ is solved to obtain $\bm{K}_\text{0}^\text{SR}$. 
The eigenstates of $\hat{H}_\text{as} =\hat{H}_\text{0,as}+\hat{H}_\text{fs}$ 
\color{black}
are well approximated by the TAM basis functions $|(NS)jlJM\rangle$ for $^{2S+1}\Sigma$ molecules \cite{Corey_83} provided the weak spin-spin interaction between the different $N$ states can be neglected.
Our FT  employs an analytical recoupling formula between the TRAM and TAM bases (see Eq.~(2.27) of Ref.~\cite{Corey_83} and the Supplemental Material \cite{SM}), which has been successfully applied to calculate the $T$-matrix for atom-molecule  collisions above 5~K \cite{Corey_83,Alexander_85,Corey_85,Lique_11,Lique_19}.  

%%%%%%%%%%%%%%%%%%%%%%%%%%%%%%%%%%%%%%%%%%%%%%%%%%%%%%
%\begin{figure}[b!]
\begin{figure}[t!]
\begin{center}
\includegraphics[height=0.17 \textheight, trim = 35 0 0 0]{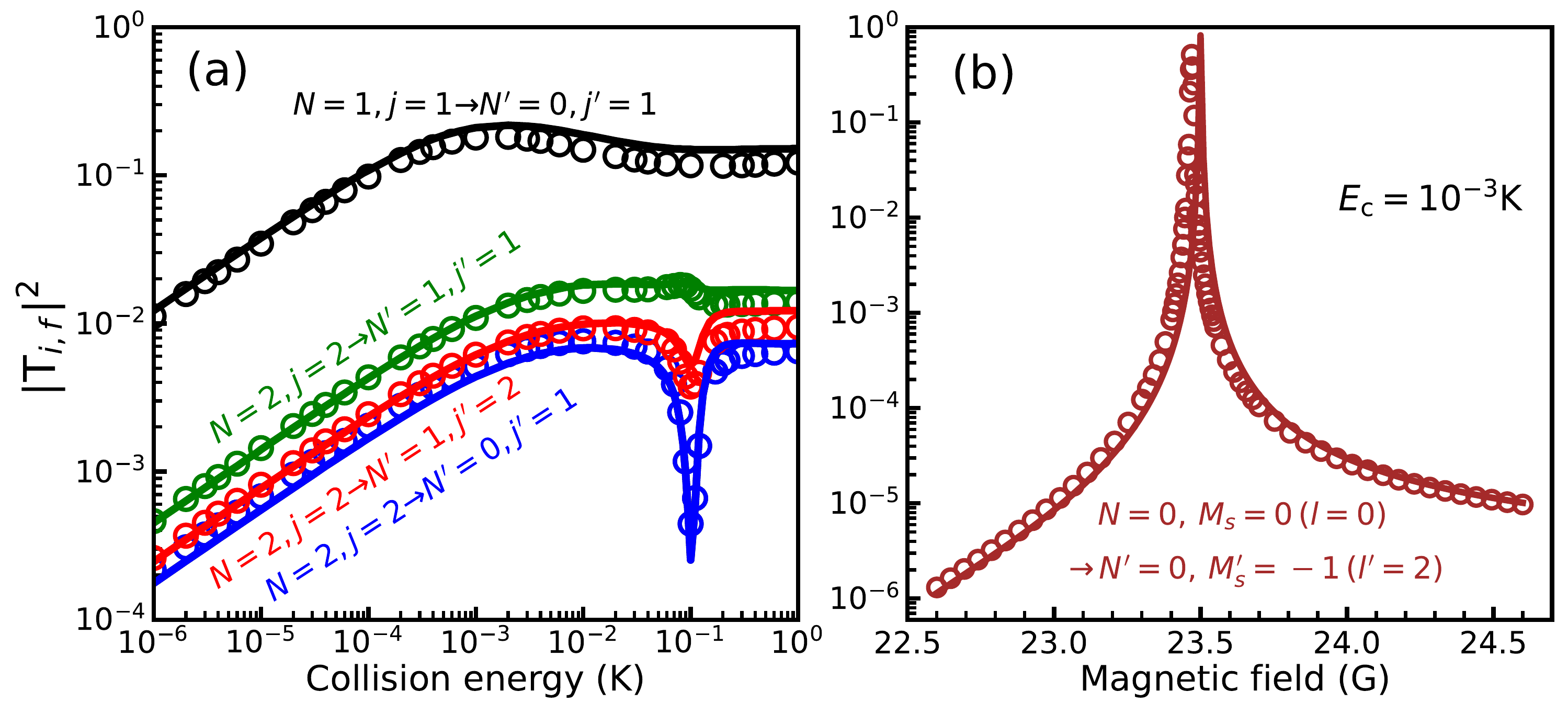}
\end{center}
\vspace{-0.2cm}
\caption{{(a)} Collision energy dependence of transition probabilities between the different fine-structure levels of $^{14}$NH in ultracold collisions with Mg atoms. %Symbols 
Solid lines -- MQDT-FT calculations, 
open circles
--  exact CC results obtained using the recoupling FT.  All results are for a single incident $s$-wave component ($l=0$) and summed over all final $l'$.
(b) Magnetic field dependence of the spin relaxation probability  in cold Mg~+~$^{14}$NH$(N=0)$ collisions. Solid line -- MQDT-FT calculations, open circle -- exact CC results. The quantum numbers for the initial and final states involved in each transition are indicated next to the corresponding curves.
% for the  $M_S=0, l=0$ $\to$  $M^\prime_S=-1, l^\prime=2$ transition
%(a) The incoming partial $s$-wave component ($l=0$) for the fine state transition from $N=1, j=1$ to $N^\prime=0, j^\prime=1$ is shown by the color of black, and the transitions from $N=2,j=2$ to $N^\prime=1,j^\prime=2$, $N^\prime=1,j^\prime=1$, and $N^\prime=0, j^\prime=1$ are shown by the color of red, green, blue. (b) The incoming partial $s$-wave component ($l=0$) for the fine state transition from $N=3, j=3$ to $N^\prime=3,j^\prime=4$ (magenta), $N^\prime=2,j^\prime=2$ (brown), and $N^\prime=1,j^\prime=2$ (light blue).
}
\label{Fig_4}
\end{figure}
%%%%%%%%%%%%%%%%%%%%%%%%%%%%%%%%%%%%%%%%%%%%%%%%%%%%%%

%\color{red} 
%Therefore, the analytical recoupling formula between $|(Nl)J_rM_r\rangle$  and $|(NS)jlJM\rangle$  \cite{Corey_83,SM}
%can be viewed as a frame transformation between the basis sets of $|(Nl)J_rM_r\rangle |SM_S\rangle$ and $|(NS)jlJM\rangle$.

%To obtain the desired short-range $K$-matrix in
 %the $|(NS)jlJM\rangle$ basis from that in the  $|(Nl)J_rM_r\rangle$ basis, we use an analytical recoupling transformation \cite{Corey_83,SM}.
%\begin{equation}
%\begin{split}
%K^{\text{sr},J}_{N'Sj'l',NSjl} =  \sum_{J_r=|N-l|}^{N+l} [J_r][j' j]^{1/2} (-1)^{-l'+l+j'-j}   
%begin{Bmatrix} S & N' & j' \\ l' & J & J_r \end{Bmatrix} 
%\begin{Bmatrix} S & N  & j  \\ l & J & J_r \end{Bmatrix}
%K^{\text{sr},J_r}_{0,N'l',Nl},
%\label{eq:7}
%\end{split}
%\end{equation}
%where $\{\}$ denotes a 6-$j$ symbol and $[X]=(2X+1)$.  

\Cref{Fig_4}(a)
shows that the probabilities for rotationally inelastic fine-structure transitions in ultracold Mg~+~NH collisions are well described by MQDT-FT. %based on the analytic FT matrix  (see also \cite{SM}). 
Thus, using a spin-independent $\hat{H}_0$ in short-range CC calculations in the framework of MQDT-FT enables one to describe a wide range of transitions between rotational and fine-structure levels of open-shell molecules in ultracold collisions with atoms.

%it is sufficient to use describe  ultracold atom-molecule collision dynamics at short range.

%approximated with recoupling technique
%. The fine state transitions accompanied by the rotational transitions
% squares of the modulus of the $T$-matrix elements for the final state resolved fine state transitions for the incoming s-wave scattering
%In particular, we observe the correct resonance feature in the transitions from the $N=2,j=2$ state in (a). We find an excellent agreement for a fine state transition without changing the rotational state $N=3$ (magenta in (b)). However, it is not necessarily possible to obtain such good agreement for all the transitions without being accompanied by the rotational transitions \cite{SM}.    
%open-shell molecular collisions

We finally subject our MQDT-FT approach to a stringent test by exploring its ability to reproduce magnetic Feshbach resonances in atom-molecule collisions, whose positions and widths are highly sensitive to fine details of short-range atom-molecule interactions \cite{Morita:19b}.
%  providing a sensitive probe of how well the short-range collision physics is captured by MQDT-FT.
%Finally, we demonstrate the efficacy  in describing magnetically tunable Feshbach resonances (see also SM \cite{SM}).
 %for $M_S=0$ ($l=0$) $\to$  $M^\prime_S=-1$ ($l^\prime=2$) in cold collision of 
In \cref{Fig_4}(b) we show the magnetic field dependence of the spin-relaxation probability in cold Mg~+~$^{14}$NH$(N=0)$ collisions, which displays a pronounced Feshbach resonance  due to  closed-channel states in the $N=0$ and $N=1$ manifolds coupled to the incident channel by the short-range anisotropic Mg-NH interaction.
% and the intramolecular interaction in NH. 
We observe excellent agreement between MQDT-FT calculations and exact CC results,
 demonstrating that the MQDT-FT approach is capable of predicting the properties of magnetic Feshbach resonances. 
The error in the resonance position does not exceed 0.1\%, comparable to the performance of MQDT-FT for ultracold Rb~+~Rb collisions  \cite{Burke_98}.
%both the position and width of the magnetic Feshbach resonance calculated using  our MQDT-FT method are in very good agreement with exact CC calculations. This demonstrates the ability of our MQDT-FT approach to predict the properties of magnetic Feshbach resonances.
%The error in the resonance position does not exceed 0.1\%, comparable to the performance of MQDT-FT for ultracold alkali-atom collisions \cite{Burke_98}.
\color{black}
%coupling between the incident collision channel with

In summary, we have generalized the powerful MQDT-FT approach to ultracold atom-molecule collisions in external magnetic fields, providing a robust conceptual and numerical framework for their theoretical description.
We have applied the approach to a realistic atom-molecule collision system (Mg~+~NH) using a variety of short-range Hamiltonians, obtaining encouraging agreement with exact CC calculations in all cases. 
Our calculations show that MQDT-FT can provide a dramatic ($10^4$-fold) reduction of computational effort of CC calculations of atom-molecule collisions compared to standard MQDT \cite{Croft_12}.
% due to the fact that there is no need to explicitly account for the hyperfine-Zeeman structure of the collision partners at short range. 
%\textcolor{red}{Efficiency of MQDT-FT becomes exponentially prominent with increasing the number of channels. }
In many cases of practical interest, it will only be necessary to solve CC equations in the strong interaction region employing a rovibrational (TAM) basis in the absence of external fields,  which can be accomplished using currently available computational techniques \cite{ABC,Pack_87,Kendrick_18,Croft_17,Morita_23,Jisha_14}. This opens up the possibility of performing rigorous quantum scattering calculations on a wide array of previously intractable ultracold atom-molecule collisions and chemical reactions in external fields, including those probed in recent experiments \cite{Yang_19,Wang_21,Son_22,Park:23b,Nichols_22}.
%which result from not having to include the spin degrees of freedom and the associated fine, hyperfine, and Zeeman interactions of the colliding molecules at short range.  

From a conceptual viewpoint, our results show that it is possible to describe the intricate quantum dynamics of ultracold atom-molecule collisions over wide ranges of collision energies and magnetic fields in terms of a small number of short-range parameters (three for Mg+NH). This provides a solid basis for the development of few-parameter MQDT-FT models of ultracold atom-molecule (and possibly even molecule-molecule \cite{Park_23}) collisions. 
%Such models, which have been so fruitful in the field of ultracold atomic collisions \cite{Burke_99,Gao_05}, may allow for novel insights into fascinating quantum dynamics of ultracold molecular collisions. 

We thank Chris Greene, John Bohn, James Croft, and Brian Kendrick for discussions. 
This work was supported by the NSF CAREER program (PHY-2045681) and by the U.S. AFOSR under Contract FA9550-22-1-0361 to P.B. and T.V.T.
Computations were performed on the Niagara supercomputer at the SciNet HPC Consortium.
%and the Digital Research Alliance of Canada.
%SciNet is funded by: the Canada Foundation for Innovation; the Government of Ontario; Ontario Research Fund - Research Excellence; and the University of Toronto. We also thank the Digital Research Alliance of Canada.

\newpage

\widetext
\begin{center}
\textbf{\large Supplemental Material}

\end{center}
%%%%%%%%%% Merge with supplemental materials %%%%%%%%%%
%%%%%%%%%% Prefix a "S" to all equations, figures, tables and reset the counter %%%%%%%%%%
\setcounter{equation}{0}
\setcounter{figure}{0}
\setcounter{table}{0}

\tableofcontents

\vspace{0.8cm}
In this Supplemental Material (SM), we provide technical details of our MQDT-FT calculations as well as some additional results, which complement the figures and discussions in the main text. 
We describe the computational details in \cref{sec:SM_Computation}.
\cref{subsec:SM_MQDT_A} presents additional background information about the standard MQDT approach for atom-molecule collisions, whereas the details of the frame transformation are discussed in \cref{subsec:SM_FT}.  
\cref{sec:SM_SR,sec:SM_hf} present
the probabilities  for the  Zeeman and hyperfine transitions in cold Mg+NH($N=0$) collisions. 
Results for the short-range $K$-matrix as a function of collision energy are presented in \cref{sec:SM_Y}. 
In \cref{sec:SM_fine}, the capabilities of the recoupling FT approach are illustrated %to calculate the short-range $K$-matrix are illustrated %by MQDT-FT calculations of 
for fine structure transitions from the $N=3$, $j=3$  initial state of NH. 
Finally, the details concerning the calculation of magnetic Feshbach resonances  in cold Mg~+~NH collisions are described in \cref{sec:SM_FR}.

\setcounter{equation}{0}
\setcounter{figure}{0}

\vspace{0.5cm}
\section{\label{sec:SM_Computation} Computational details}

\subsection{Coupled channel (CC) calculations}

We performed coupled-channel (CC) calculations to obtain the reference results for cold Mg~+~NH collisions  in a magnetic field, against which we compare our MQDT-FT calculations in the main text.  
The Hamiltonian of the atom-molecule collision complex is given by Eq.~(1) of the main text.
% for the calculations of Fig.~2, by $\hat{H} - \hat{H}_\text{Z}$ for those of  Figs.~1 and 3, and by $\hat{H} - \hat{H}_\text{Z}-\hat{H}_\text{hfs}$ for those of Fig.~4.
We use converged uncoupled basis sets, $|NM_N\rangle|SM_S\rangle|I_{1} M_{I_1}\rangle|I_{2}M_{I_2}\rangle|lm_l\rangle$ for Fig.~2 and $|NM_N\rangle|SM_S\rangle|lm_l\rangle$ for Figs.~1 and 3, with $N_\text{max}=6$ and $l_\text{max}=8$ \cite{Wallis_09}. 
On the other hand, for Fig.~4, we employ a coupled total angular momentum (TAM) basis set $|(NS)jlJM\rangle$ with $N_\text{max}=6$ and $J_\text{max}=5$.
The CC equations are solved with the log-derivative propagation method \cite{Johnson_73} between $R_\text{min}= 4\,a_0$ and $R_\text{max}= 500\,a_0$. We use the propagation interval of $\Delta\,R= 0.05\,a_0$ for $R\leq30\,a_0$ and $\Delta\,R=0.1\,a_0$ for $R>30\,a_0$. 

\subsection{Physical constants}
We use the same reduced mass of Mg+$^{14}$NH ($\mu=9.232679959$ amu), and spectroscopic constants of $^{14}$NH  ($B_e=16.343$ cm$^{-1}$, $\gamma_\mathrm{sr}=-0.055$ cm$^{-1}$, and $\lambda_\mathrm{ss}=0.92$ cm$^{-1}$) as in Refs.~\cite{Wallis_09,MOLSCAT}. For the calculations  shown in Fig.~2 of the main text, we use the reduced mass of $\mu=9.60046$ amu for Mg+$^{15}$NH, the spectroscopic constants of $^{15}$NH ($B_e=16.2712$ cm$^{-1}$, $\gamma_\mathrm{sr}=-0.0546$ cm$^{-1}$, and $\lambda_\mathrm{ss}=0.9199$ cm$^{-1}$), and 
%in Ref.~\cite{Bizzocchi_18}. 
the hyperfine constants of $^{15}$NH ($b_1=-1.9406\times 10^{-3}$ cm$^{-1}$, $c_1=3.1783\times 10^{-3}$ cm$^{-1}$, $b_2=-3.2108\times 10^{-3}$ cm$^{-1}$ and $c_2=3.0203\times 10^{-3}$ cm$^{-1}$) ~\cite{Bizzocchi_18}. 
%Here, the constants $b_i$ ($i=1,2$) are obtained from the values of $b_{\text{F},i}$ and $c_i$ given in Ref.~\cite{Bizzocchi_18} with the relation $b_{\text{F},i}=b_i+c_i/3$, where $b_{\text{F},i}$ is the constant for the Fermi contact interactions. 

\subsection{Interaction potential}
The interaction potential between Mg ($^1$S) and NH ($\tilde{X}\, ^3\Sigma^-$) used in our calculations is published as part of the MOLSCAT program suite \cite{MOLSCAT} as pot-Mg\_NH.data and vstar-Mg\_NH.f. 
This potential is slightly different from that used in Refs.~\cite{Wallis_09,Maykel_11,Croft_11}.

\section{\label{sec:SM_MQDT-FT} MQDT-FT}

\subsection{\label{subsec:SM_MQDT_A} Multi-channel quantum defect theory (MQDT)}
%\subsection{Multi-channel quantum defect theory (MQDT)}
%\textcolor{red}{Except for the frame transformation step to obtain the short-range $K$-matrix ($\bm{K}^\text{sr}$),} 
%Except for the frame transformation step to approximate the short-range $K$-matrix $\bm{K}^\text{sr}$ based on $\bm{K}_\mathrm{0}^\text{sr}$ (see details in \cref{sec:SM_FT} and main text), 

Here, we provide a brief overview of the standard MQDT approach for atom-molecule collisions \cite{Croft_11,Croft_12}, focusing on the details essential for MQDT-FT.
The matrix solution $\bm{\Psi}$ of the Schr\"odinger equation $\hat{H}\bm{\Psi}=E\bm{\Psi}$ in the basis of the eigenvectors of the asymptotic Hamiltonian ($\hat{H}_\text{as}$)  is matched to the boundary conditions \cite{Mies_00,Croft_11} 
\begin{equation}
\label{eq:Boundary}
\bm{\Psi}( R_{m} )= R_{m}^{-1}\,[\bm{f}(R_{m})+\bm{g}(R_{m})\,\bm{K}^{\text{sr}}\,],
\end{equation}
%where the $N\times N$ matrix $\bm{\Psi}$ of the total wave function for the Schr\"{o}dinger equation with $\hat{\mathcal{H}}_\mathrm{0}$, and reference functions obtained by solving a set 
where the matching radius $R=R_m$ marks the boundary between the short-range and long-range regions, and $\bm{f}(R_m)$ and $\bm{g}(R_m)$ are diagonal matrices composed of the values of regular and irregular reference functions at $R_m$ for each channel \cite{Mies_84,Croft_11}. 
To obtain the reference functions, we numerically solve the one-dimensional radial Schr\"odinger equations
\begin{equation}
\label{eq:1Deq}
\left[\, - \frac{1}{2\mu} \frac{\partial^2}{\partial R^2} + U_i^\text{ref}(R)\, \right]\, \psi_i(R)\,=\, E\,\psi_i(R),
\end{equation}
where $U_i^\text{ref}(R)\, =\, V_0(R) + {l_i(l_i+1)}/{(2\mu R^2)} + E_i^{\infty}$ is the reference potential for the $i$-th channel \cite{Croft_11} with threshold energy $E_i^{\infty}$ and orbital angular momentum $l_i$, and $V_0(R)$  is the isotropic part of the interaction potential between Mg and NH.

%\vspace{0.3cm}
%\color{blue}
We note that the matrices $\bm{K}^\text{sr}$, $\bm{f}(R_m)$, $\bm{g}(R_m)$, and $\bm{\Psi}(R_m)$ have dimensions equal to the total number of reference channels $N_\mathrm{ref} = N_\mathrm{o} +N_\mathrm{wc}$ \cite{Croft_11}, where $N_\mathrm{o}$ and $N_\mathrm{wc}$ are the numbers of open and weakly closed channels, respectively.
We neglect the strongly closed channels, whose coupling with open and weakly closed channels has no effect on scattering dynamics \cite{Croft_11,Croft_12}.
Here, we define strongly closed channels as those, for which $U_i^\text{ref}(R)>E$ for all $R$.
%Once we obtain an approximation of $\bm{K}^\text{sr}$ from $\bm{K}_\mathrm{0}^\text{sr}$ by the frame-transformation (FT) (see  \cref{sec:SM_FT} and main text for the details)

The conventional MQDT procedure then proceeds to obtain the  $T$-matrix from the short-range $K$-matrix ($\bm{K}^\text{sr}$). First, the dimension of $\bm{K}^\text{sr}$ 
($N_\mathrm{ref} \times N_\mathrm{ref}$)
is reduced by taking into account the coupling between the open and weakly closed channels. This is accomplished by forming the 
($N_\mathrm{o} \times N_\mathrm{o}$)
matrix
%, $\bar{\bm{K}}^\text{sr}$,
%extract the open-open block of the short-range $K$-matrix by taking
%The $N_\mathrm{o} \times N_\mathrm{o}$ ($N_\mathrm{o}$: Number of open channels) matrix $\bar{\bm{K}}^\text{SR}$ 
\begin{equation}
\bar{\bm{K}}^\text{sr}=\bm{K}_\mathrm{o,o}^\text{sr}-\bm{K}_\mathrm{o,wc}^\text{sr}\,[\,\tan{\bm{\nu}}+\bm{K}_\mathrm{wc,wc}^\text{sr} \,]^{-1}\,\bm{K}_\mathrm{wc,o}^\text{sr},
\label{eq:Ksr_exact}
\end{equation}
%whose diagonal elements are
where $\tan{\bm{\nu}}$ is a diagonal matrix of QD parameters $\tan \nu_i$ for weakly closed channels \cite{Mies_84,Mies_00,Croft_11}.
The zeros of the bound-state phase ($\tan \nu_i=0$) correspond to the bound state energies in the $i$-th reference potential and the matrix $\bm{K}_\mathrm{wc,wc}^\text{sr}$ introduces shifts in resonance positions \cite{Mies_00}.  

In the next step, threshold effects are taken into account using the diagonal matrices of QD parameters, $\bm{C}$ and $\tan{\bm{\lambda}}$, for open channels to obtain the $\bar{\bm{R}}$ matrix 
\begin{equation}
\bar{\bm{R}}=\bm{C}^{-1}\,[\,({\bar{\bm{K}}^\text{sr}})^{-1}-\tan{\bm{\lambda}} \,]^{-1}\,\bm{C}^{-1},
\label{eq:Rbar}
\end{equation}
where $C_i$ determines the amplitude of the reference functions and $\lambda_i$ is the phase difference between the long-range and short-range irregular solutions \cite{Raoult_04}.
%(the deviation of $\lambda_i$ from 0 is a manifestation of the breakdown of the WKB approximation
Finally, the  transition $T$-matrix may be written as 
\begin{equation}
\bm{T}=\bm{I}-e^{i\bm{\xi}}\,[\,\bm{I}+i\bar{\bm{R}} \,]\,[\,\bm{I} - i\bar{\bm{R}} \,]^{-1}\,e^{i\bm{\xi}},
\label{eq:Smatrix}
\end{equation}
where $\bm{I}$ is the unit matrix and $\bm{\xi}$ is the diagonal matrix of phase shifts $\xi_i$ associated with the energy-normalized reference functions  \cite{Mies_84,Mies_00,Raoult_04,Croft_11,Croft_12}. The state-to-state integral cross sections are 
obtained from the transition probability $|T_{i,f}|^2$ as
$\sigma_{i \to f}=(\pi/k_i^2)|T_{i,f}|^2$, where $i$ and $f$ label the initial and final internal molecular states, $k_i=\sqrt{2\mu E_\mathrm{c}}$ is the incident wavevector, and $E_\mathrm{c}$ is the collision energy.

\subsection{\label{subsec:SM_FT} Frame transformation (FT)}
%\textcolor{red}{
%The selection of $H_0$ is not straightforward for ultracold molecular collisions. Previous applications of MQDT-FT to the collisions of ultracold alkali atoms relied on the assumption that spin-changing processes are primarily driven by spin-exchange interactions characterized by the difference of the singlet and triplet interaction potentials at short-range. However, recent intriguing applications of ultracold molecular collisions, such as sympathetic cooling and evaporative cooling, predominantly proceed on a single potential energy surface. Consequently, intramolecular interactions, such as spin-rotation and spin-spin interactions, become the primary mechanisms responsible for the spin-changing processes at short-range. Furthermore, the impact of intramolecular interactions varies in terms of the values of the spectroscopic constants, which may result in the non-negligible coupling between different rotational states (e.g. spin-spin interaction: $N=0 \leftrightarrow N=2$). Thus, unlike alkali atomic collisions, it is not possible to provide a general and analytical form of frame transformation even when neglecting the magnetic field dependence and specifying the quantum numbers of spins in the system. 
%}

{
Here, as in the main text, $\hat{H}$ denotes the Hamiltonian of the atom-molecule collision complex, and $\hat{H}_0$ denotes the simplified Hamiltonian, {from which the short-range $K$-matrix, $\bm{K}_\mathrm{0}^\text{sr}$, is obtained using the  MQDT-FT boundary conditions given by Eq.~(4) of the main text.}
%\color{red}
%determing $\bm{K}_\mathrm{0}^\text{sr}$. 
%\color{black}
%in short-range CC calculations of $\bm{K}_\mathrm{0}^\text{sr}$ in MQDT-FT. 
We approximate $\bm{K}^\text{sr}$ by the matrix obtained via the frame transformation (FT) as $\bm{K}^\text{sr} \simeq \bm{U}^{\dagger}\bm{K}_\mathrm{0}^{\text{sr}}\bm{U}$, where %$\bm{K}_\mathrm{0}^{\text{sr}}$ is the short-range $K$-matrix with $\hat{H}_0$ and 
$\bm{U}$ is a unitary matrix composed of the eigenstates of the asymptotic Hamiltonian $\hat{H}_\text{as}=\lim_{R\to \infty} \hat{H}$.
Since $\bm{K}_\mathrm{0}^{\text{sr}}$ is initially expressed in the basis of eigenstates of the asymptotic Hamiltonian $\hat{H}_{0,\text{as}} = \lim_{R\to \infty}\hat{H}_\mathrm{0}$, it is necessary to establish a relationship between the eigenstates of $\hat{H}_\text{as}$ and $\hat{H}_{0,\text{as}}$ to represent $\bm{K}^\text{sr}$ in the basis of eigenstates of  $\hat{H}_\text{as}$.
}

{Having obtained the short-range $K$-matrix, we follow the standard MQDT steps as described below.}
 First, it is necessary to obtain reference functions and associated quantum defect (QD) parameters by solving \cref{eq:1Deq}.  {In doing so, we use the accurate threshold energies $E_i^{\infty}$ obtained by diagonalizing the exact asymptotic Hamiltonian $\hat{H}_\text{as}$.  } 
 The reference potentials contain a hard wall at $R=R_\text{wall}$ in the short range region ($U_i^\text{ref}(R_\text{wall})=+\infty$) restricting the amplitude of the regular reference function at $R=R_\text{wall}$.
We employ a common wall position $R_\text{wall}=5\,a_\text{0}$ for all channels although the final results are not affected by the choice of $R_\mathrm{wall}$. 
Here, we use the reference potentials with minima at a finite $R$ in the range $0<R<R_m$ and the WKB normalizations are taken at the minimum of the reference potentials to obtain the reference functions $f_i(R)$ and $
g_i(R)$, following Mies  {\it el al.} \cite{Mies_84}. 

The value of the matching radius $R_m$ should be determined carefully by exploring the $ R_m$ dependence of the results.
We use $R_m=30\, a_0$ for all MQDT-FT calculations presented  here and in the main text. No sensitivity to $R_m$ was observed in the range  $15\, a_0 \leq R_m \leq 40\, a_0$ for the results shown in Fig.~1 of the main text and in the range $25\, a_0 \leq R_m \leq 35\, a_0$  for those shown in Fig.~2 of the main text.

%represented in the eigenstates of $\hat{H}_\text{as}$ for the succeeding steps in the MQDT-FT calculation, it is necessary to know the relation between the eigenstates of $\hat{H}^\text{as}$ and $\hat{H}_0^\text{as}$. \\
%$\bm{U}$ (and $\hat{H}^\text{as}$) with the eigenstates of $\hat{H}_0^\text{as}$. Depending on the technical convenience and available basis sets, we can perform FT using different representations. For example, when we obtain $\bm{U}$, eigenvectors of $\hat{H}^\text{as}$, by a basis set, it is a way to represent $\bm{K}_\mathrm{0}^{\text{sr}}$ in the basis set to perform FT ($\bm{K}^\text{sr} \simeq \bm{U}^{\dagger}\bm{K}_\mathrm{0}^{\text{sr}}\bm{U}$).  
%If $\hat{H}_0^\text{as}$ does not include some (spin) degrees of freedom compared to $\hat{H}^\text{as}$, like the nuclear spins in the calculations of Fig.~2 and \cref{SM_Fig_2}, we need to augment the missing degrees of freedom to the basis set or eigenstates of $\hat{H}_0^\text{as}$. 
%Such transformation is possible only when the contained degrees of freedom are equal.  %into the representation of the 
%represent $\bm{U}$ with eigenstates of $\hat{H}_0^\text{as}$ or represent $\bm{K}_\mathrm{0}^{\text{sr}}$ with the basis set used in the representation of $\bm{U}$.
%\vspace{0.2cm}

%While the above procedure is general, we can consider efficient variants of FT based on the character of the relevant channels in the collisions. 
% as a possible technical difficulty in molecular collisions, 
% and thus not all channel basis states need to be c in MQDT-FT calculations.  

Once we represent $\hat{H}_{0,\text{as}}$ and  $\hat{H}_\text{as}$  
in the same basis set, it is straightforward to apply the FT. However, it can be challenging to obtain the eigenvectors of $\hat{H}_\text{as}$ in the basis set used in short-range CC calculations
with $\hat{H}_0$. The latter are best performed using the TAM basis set, and it is not straightforward to derive and/or implement the matrix elements of $\hat{H}_\text{as}$ in this basis. 
To avoid these difficulties, % we pre-transform $\bm{K}_\mathrm{0}^{\text{sr}}$ to 
we propose to use the fully uncoupled basis set, in which the matrix elements of $\hat{H}_\text{as}$ are readily available \cite{Krems_04}.

%We note that the actual dimension of $\bm{K}_\mathrm{0}^\text{sr}$, $\bm{f}(R_m)$, $\bm{g}(R_m)$, and $\bm{\Psi}(R_m)$ is equal to the total number of reference channels $N_\mathrm{ref} = N_\mathrm{o} +N_\mathrm{wc}$ \cite{Croft_11}, where $N_\mathrm{o}$ and $N_\mathrm{wc}$ are the numbers of open and weakly closed channels, respectively.
%We neglect the strongly closed channels, whose coupling with open and weakly closed channels has no effect on scattering dynamics \cite{Croft_11,Croft_12}.
%Here, we define strongly closed channels as those, for which $U_i^\text{ref}(R)>E$ for all $R$.

% in the fully uncoupled basis. 
%We note that strongly closed channels are excluded from the calculation of  $\bm{K}_\mathrm{0}^{\text{sr}}$ (see above).

For example, to obtain the results  shown in Fig.~1 of  the main text, the following sequence of steps is used

\begin{enumerate}

\item
{
The matrix of $\hat{H}_\text{as}$ is constructed and diagonalized in the fully uncoupled basis  $|NM_N\rangle|SM_S\rangle|lm_l\rangle$. 
The resulting eigenstates are expressed as linear combinations of the fully uncoupled basis functions
\begin{equation}
\label{eq:eigenstate}
|\, \alpha\, \rangle\, =\,  \sum_{i}^{}\, c^\alpha_i |\, i\, \rangle,
\end{equation}
where $|\,i\,\rangle$ is a short-hand notation for the uncoupled basis functions $|NM_N\rangle|SM_S\rangle|lm_l\rangle$, which includes the quantum numbers $N$, $M_N$, $M_S$, $l$, and $m_l$ (note that $S=1$ is conserved), and the expansion coefficients \{$c_i^\alpha$\} correspond to the $\alpha$-th eigenvector of $\hat{H}_\text{as}$.  
The calculated eigenvectors are stored for subsequent use in step 4. 
}

\item
{
The CC equations for the simplified Hamiltonian $\hat{H}_0$ expressed in the primitive TAM basis  $|(NS)jlJM\rangle$ are integrated numerically out to $R=R_m$.  
The log-derivative matrix is then transformed from the primitive TAM basis set to the basis of eigenvectors of $\hat{H}_{0,\text{as}}$ to obtain
the short-range $K$-matrix $\bm{K}_\mathrm{0}^{\text{sr}, J}$ using the matching condition given by %\cref{eq:Boundary} 
Eq.~(4) of the main text
for each $J$ ($J_\text{max}=5$). 
%We note that components associated with strongly closed channels are excluded from $\bm{K}_\mathrm{0}^{\text{sr},J}$ (see above).
Thus, $\bm{K}_\mathrm{0}^\text{sr}=\bm{K}_\mathrm{0}^{\text{sr}, J=0} \oplus \bm{K}_\mathrm{0}^{\text{sr},J=1} \oplus \cdots \oplus \bm{K}_\mathrm{0}^{\text{sr},J=5}$.
}

%\textcolor{magenta}{
%We obtain 
%$\bm{K}_\mathrm{0}^{\text{sr}}$ as $\bm{K}_\mathrm{0}^\text{sr}=\bm{K}_\mathrm{0}^{\text{sr}, J=0} \oplus \bm{K}_\mathrm{0}^{\text{sr},J=1} \oplus \cdots \oplus \bm{K}_\mathrm{0}^{\text{sr},J=6}$.  
%}

\item
{
For Mg~+~NH collisions, the eigenstates of $\hat{H}_{0,\text{as}}$ are well approximated by the individual TAM basis functions due to the small spin-spin interaction in NH (in other words, the matrix of $\hat{H}_{0,\text{as}}$ is nearly diagonal in the TAM basis). Therefore, in the following, we assume that $\bm{K}_\mathrm{0}^{\text{sr}}$ is expressed in the TAM basis. If $\hat{H}_{0,\text{as}}$ is not diagonal in this basis,  $\bm{K}^\text{sr}$ would need to be transformed to the basis, in which  $\hat{H}_{0,\text{as}}$ is diagonal (see the discussion after Step 4).
}

\vspace{0.2cm}
{
The matrix elements of $\bm{K}^\text{sr}$ in the basis of eigenstates $\alpha$ and $\beta$ of $\hat{H}_{\text{as}}$ is obtained by applying the FT
\begin{equation}
\label{eq:FT}
\langle\, \alpha\, |\, \bm{K}^\text{sr}\, |\, \beta\, \rangle =  \sum_{p,q}^{}\, {d_p^\alpha}^\ast d_q^\beta \langle\, p\, |\, \bm{K}_0^\text{sr}\, |\, q\, \rangle,
\end{equation}
where $|p\rangle$ and $|q\rangle$ are the TAM basis functions, \{$d_p^\alpha$\} and \{$d_q^\beta$\} are the eigenvectors of $\hat{H}_{\text{as}}$ in the TAM basis. Instead of calculating these eigenvectors directly, we express them via the eigenvectors of $\hat{H}_{\text{as}}$ in the fully uncoupled basis (see Step 1) as described below. 
%an alternative method based on the eigenvectors of step 1. 
}

\item

As noted above, we wish to calculate $\bm{K}^\text{sr}$ in the eigenbasis of $\hat{H}_{\text{as}}$ with the latter operator is expressed in the fully uncoupled basis. This is motivated by the ease of evaluating  the matrix elements of $\hat{H}_{\text{as}}$ in the fully uncoupled basis \cite{Krems_04}.   
It follows from \cref{eq:eigenstate} that 
\begin{equation}
\label{eq:uncFT}
\langle\, \alpha\, |\, \bm{K}^\text{sr}\, |\, \beta\, \rangle =  \sum_{i,j}^{}\, {c_i^\alpha}^\ast c_j^\beta \langle\, i\, |\, \bm{K}_0^\text{sr}\, |\, j\, \rangle.
\end{equation}

To express the matrix elements on the right-hand side in terms of the known quantities $\langle\, p\, |\bm{K}_0^\text{sr}\,|\, q\, \rangle$, we use the completeness relation $\hat{1}=\sum_p |p\rangle \langle p|$:
%However, the transformation of the basis of $\bm{K}_0^\text{sr}$ is simple, thus we evaluate the following relation to obtain $\bm{K}^\text{sr}$
\begin{equation}
\label{eq:finalFT}
\langle\, \alpha\, |\, \bm{K}^\text{sr}\, |\, \beta\, \rangle =  \sum_{i,j}^{}\,{c_i^\alpha}^\ast c_j^\beta \langle\, i\, |\,  \left[\, \sum_{p,q}^{} |\,p\,\rangle\,\langle\, p\, |\bm{K}_0^\text{sr}\,|\, q\, \rangle\,\langle\, q\, |\, \right]\ |\, j\, \rangle \\
 =  \sum_{i,j,p,q}^{}\, {c_i^\alpha}^\ast c_j^\beta \langle\, i\, |\,p\,\rangle\, \langle\, q\, |\,j\,\rangle\,  \langle\, p\, |\bm{K}_0^\text{sr}\,|\, q\, \rangle\, ,
\end{equation}
where the overlaps between the fully uncoupled basis state $|\,i\,\rangle$ and the TAM basis state $|\,p\,\rangle$ are given analytically by Clebsh-Gordan coefficients. We note that the expression in \cref{eq:finalFT} is more generally  applicable to the cases, where the  overlaps $\langle\, i\, |\,p\,\rangle$ are only available from a numerical computation (which could rely on approximations). Additionally, it is not necessary to explicitly construct the matrix $\bm{K}_0^\text{sr}$ in the fully uncoupled basis set in this calculation.   

\end{enumerate}

\vspace{0.3cm}
For completeness, we also consider the situation, in which the eigenstates of $\hat{H}_{0,\text{as}}$ are strongly different from the individual TAM basis functions (this is not the case in the present work). In this situation, the assumption discussed in Step 3 above does not hold, and additional summations over the TAM basis functions and the associated overlap coefficients (eigenvectors) appear in \cref{eq:finalFT}. Let $|\phi\rangle$ be an eigenstate of $\hat{H}_{0,\text{as}}$, which is expanded in the TAM basis set as 
\begin{equation}
\label{eq:eigenstateH0as}
|\, \phi\, \rangle\, =\,  \sum_{p}^{}\, h_p^\phi |\, p \, \rangle.
\end{equation}
From \cref{eq:finalFT}, we obtain 
\begin{equation}
\label{eq:finalFT2}
\langle\, \alpha\, |\, \bm{K}^\text{sr}\, |\, \beta\, \rangle 
 =  \sum_{i,j,\phi,\chi}^{}\, {c_i^\alpha}^\ast c_j^\beta \langle\, i\, |\,\phi \,\rangle\, \langle\, \chi \, |\,j\,\rangle\,  \langle\, \phi \, |\bm{K}_0^\text{sr}\,|\, \chi \, \rangle\,.
\end{equation}
The matrix elements $\langle\, \phi \, |\bm{K}_0^\text{sr}\,|\, \chi \, \rangle$ and the expansion coefficients \{$h_p$\}  \cref{eq:eigenstateH0as} are available from short-range CC calculations in the TAM basis  subject to the matching condition Eq.~(4) of the main text. %\cref{eq:Boundary}. 

The most general FT is then obtained using \cref{eq:eigenstateH0as} as  
\begin{equation}
\label{eq:finalFT3}
\langle\, \alpha\, |\, \bm{K}^\text{sr}\, |\, \beta\, \rangle 
 =  \sum_{i,j,\phi,\chi}^{}\, {c_i^\alpha}^\ast c_j^\beta 
 \langle\, \phi \, |\bm{K}_0^\text{sr}\,|\, \chi \, \rangle\
 \sum_{p,q}^{}\, 
 {h_p^\phi} {h_q^\chi}^\ast  \langle\, i\, |\,p \,\rangle\, \langle\, q \, |\,j\,\rangle\,.
\end{equation}

\vspace{0.5cm}
To produce the results shown in Fig.~2 of the main text, we implemented the following sequence of MQDT-FT steps

\begin{enumerate}

\item
{
Build and diagonalize the matrix of $\hat{H}_\text{as}$ (including the hyperfine interactions) in the fully uncoupled basis set $|NM_N\rangle|SM_S\rangle|lm_l\rangle|I_1M_{I_1}\rangle|I_2M_{I_2}\rangle$. 
}

\item
{
This step is the same as Step 2 above. The only difference is that the number of $J$-blocks is increased to $J_\text{max}=7$. 
}

\item
{
As in Step 3 above, we assume that the individual TAM basis functions $|(NS)jlJM\rangle$ are the eigenstates of $\hat{H}_{0,\text{as}}$. In this case, $\hat{H}_\text{as}$ includes the nuclear spin degrees of freedom for the two nuclei of NH but $\hat{H}_{0,\text{as}}$ is identical to that used above (see the steps for Fig.~1).

Because the short-range $K$-matrix is computed with the nuclear spin degrees of freedom omitted, we need to reintroduce them  before applying MQDT boundary conditions (which do include the nuclear spin degrees of freedom). To this end, we augment $\bm{K}_\mathrm{0}^\text{sr}\to \bm{K}_\mathrm{0}^\text{sr}\otimes\bm{I}$, where $\bm{I}$ is a $4\times4$ unit matrix in the basis of 4 possible nuclear  spin states of $^{15}$NH, $|M_{I_1} =\pm1/2,M_{I_2} =\pm1/2\rangle$. 
% $M_{I_1} =1/2$ and $M_{I_2} =-1/2$,\, $M_{I_1}=-1/2$ and $M_{I_2} =1/2$, and $M_{I_1}=M_{I_2} =-1/2$.
The new basis for $\bm{K}_\mathrm{0}^\text{sr}\otimes\bm{I}$ is given by the direct product of the TAM basis and $|I_1M_{I_1}\rangle|I_2M_{I_2}\rangle$, namely $|(NS)jlJM\rangle|I_1M_{I_1}\rangle|I_2M_{I_2}\rangle$. 
}

\item
{
Finally, we identify $|i\rangle \leftrightarrow |NM_N\rangle|SM_S\rangle|lm_l\rangle|I_1M_{I_1}\rangle|I_2M_{I_2}\rangle$ and $|p\rangle \leftrightarrow |(NS)jlJM\rangle|I_1M_{I_1}\rangle|I_2M_{I_2}\rangle$ and use \cref{eq:finalFT} to obtain the desired matrix elements of $\bm{K}^{\text{sr}}$ in the basis of eigenstates of $\hat{H}_\text{as}$.
}

\end{enumerate}

\vspace{0.2cm}
\section{\label{sec:SM_SR} Electron spin transitions}

%%%%%%%%%%%%%%%%%%%%%%%%%%%%%%%%%%%%%%%%%%%%%%%%%%%%%%
\begin{figure}[t!]
\begin{center}
\includegraphics[height=0.32
\textheight,keepaspectratio]{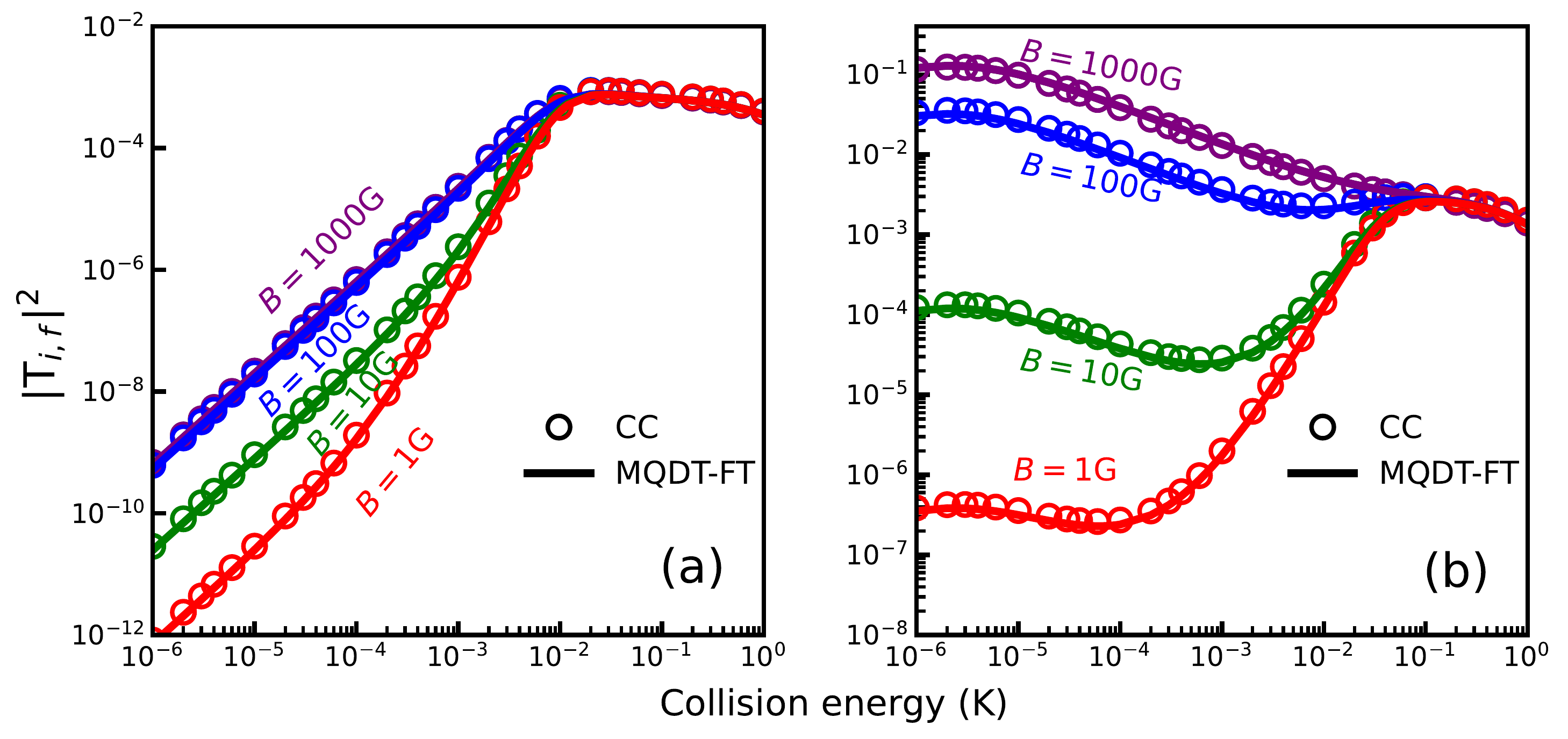}
\end{center}
\caption{Probabilities for the spin relaxation transitions (a) $M_S=1$ ($l=1$) $\to$ $M^\prime_S=0$ ($l^\prime=1$) and (b) $M_S=1$ ($l=0$) $\to$ $M^\prime_S=-1$ ($l^\prime=2$) in Mg~+~$^{14}$NH ($N=0$) collisions. Open circles --  exact CC calculations, solid lines -- MQDT-FT results. The magnitude of the external magnetic field is indicated next to each curve. 
%shown by the color of purple ($B=1000$\,G), blue ($B=100$\,G), green ($B=10$\,G), and red ($B=1$\,G).
}
\label{SM_Fig_1}
\end{figure}
%%%%%%%%%%%%%%%%%%%%%%%%%%%%%%%%%%%%%%%%%%%%%%%%%%%%%%

%%%%%%%%%%%%%%%%%%%%%%%%%%%%%%%%%%%%%%%%%%%%%%%%%%%%%
\begin{figure}[b!]
\begin{center}
\includegraphics[height=0.33\textheight,keepaspectratio]{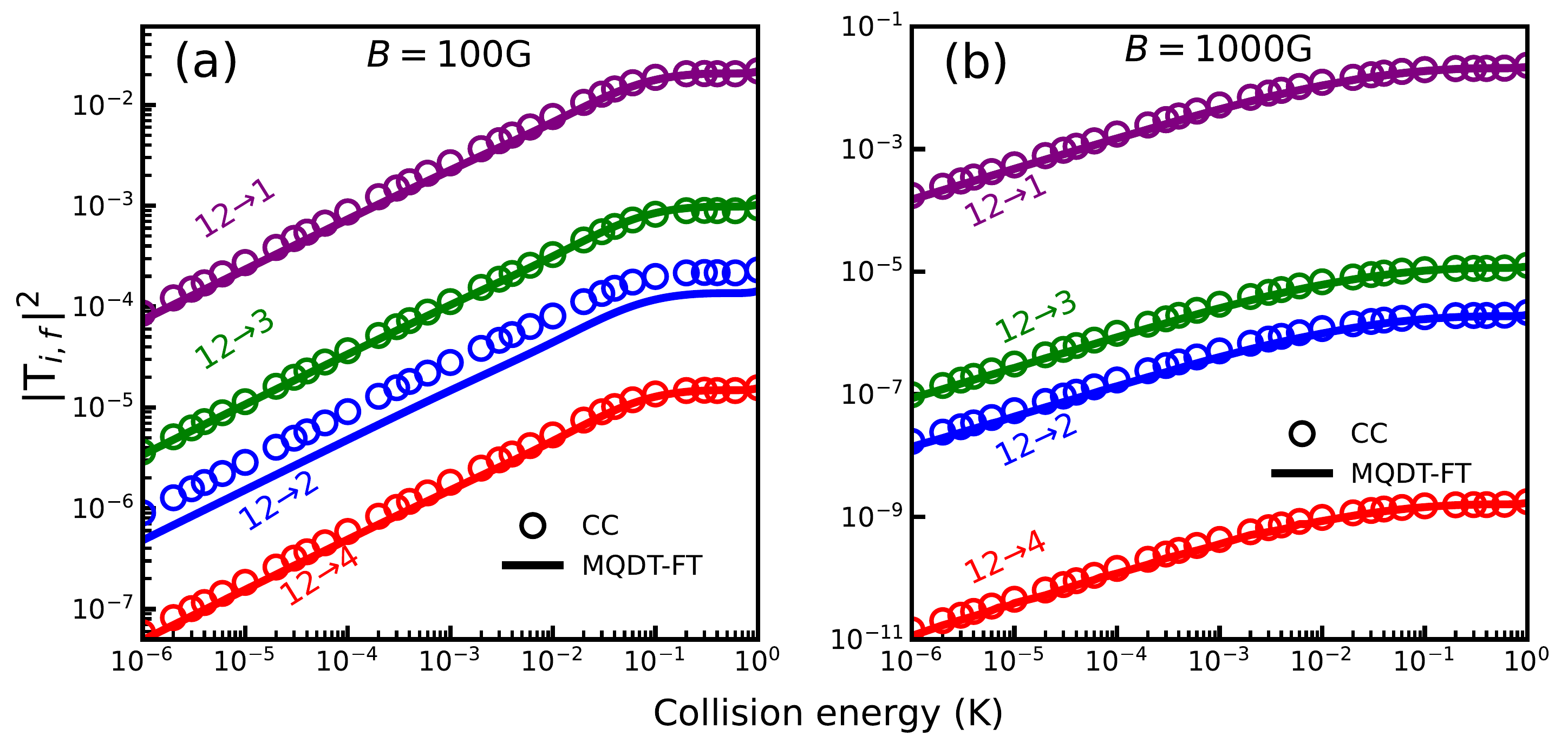}
\end{center}
\caption{
Probabilities for the state-to-state transitions ($l=0$ $\to$ $l^\prime=2$) at the magnetic field of (a) $B=100$\,G (b) and $B=1000$\ 
in Mg+$^{15}$NH ($N=0$) collisions.
Open circles --  exact CC calculations, solid lines -- MQDT-FT results. The initial and final states of $^{15}$NH ($N=0$) are indicated next to each curve.
%as $12 \to 1$ (purple), $12 \to 2$ (blue), $12 \to 3$ (green), and $12 \to 4$ (red). 
}
\label{SM_Fig_2}
\end{figure}
%%%%%%%%%%%%%%%%%%%%%%%%%%%%%%%%%%%%%%%%%%%%%%%%%%%%%%

To further illustrate the good agreement between our MQDT-FT results and exact CC calculations (see Fig.~1 of the main text), we show in \cref{SM_Fig_1} a sampling of state-to-state transitions in ultracold Mg~+~NH collisions. 
While our main interest here is in ultracold $s$-wave collisions, we also show the results for the $M_S=1$ ($l=1$) $\to$ $M^\prime_S=0$ ($l^\prime=1$) transition in panel (a) to demonstrate the capability of MQDT-FT to describe $p$-wave collisions. In panel (b), we show the results for the $M_S=1$ ($l=0$) $\to$ $M^\prime_S=-1$ ($l^\prime=2$) transition, which is similar to the  $M_S=1$ ($l=0$) $\to$ $M^\prime_S=0$ ($l^\prime=2$) transition (see Fig.~1 of the main text).  
%We show the MQDT-FT results for the $M_S=1$ ($l=0$) $\to$ $M^\prime_S=0$ ($l^\prime=2$) transition in Fig.~1 in the main text. Here we show other relevant partial-wave resolved state-to-state transitions in \cref{SM_Fig_1}. While our main concern in this paper is the ultracold (incoming) $s$-wave collisions, we show the results for $M_S=1$ ($l=1$) $\to$ $M^\prime_S=0$ ($l^\prime=1$) in (a) to demonstrate the capability of the MQDT-FT method for $p$-wave. In (b), we show $M_S=1$ ($l=0$) $\to$ $M^\prime_S=-1$ ($l^\prime=2$), which is similar to the $M_S=1$ ($l=0$) $\to$ $M^\prime_S=0$ ($l^\prime=2$) in Fig.~1 of the main text. 

%\vspace{0.5cm}
\section{\label{sec:SM_hf} Hyperfine transitions}

In \cref{SM_Fig_2}, we show the cross sections for the hyperfine transitions to the final states of $^{15}$NH ($N=0$) corresponding to $M^\prime_S=-1$ (see the main text) in $s$-wave collisions of Mg atoms with  $^{15}$NH molecules initially in the highest energy state (12). We observe excellent agreement between MQDT-FT and exact CC results for most transitions. The agreement improves at higher magnetic fields.
%initially in the highest state (12th state) of the rotational ground state of NH. We observe excellent agreement with the rigorous CC results despite of neglecting the nuclear spin-dependent Hamiltonian in the short-range for the MQDT-FT calculation, leading to a significant reduction of computational time and the amount of memory use. 
%On the other hand, we observe a slight discrepancy for the $12 \to 2$ transition.    

 %%%%%%%%%%%%%%%%%%%%%%%%%%%%%%%%%%%%%%%%%%%%%%%%%%%%%%
\begin{figure}[t!]
\begin{center}
\includegraphics[height=0.28\textheight,keepaspectratio]{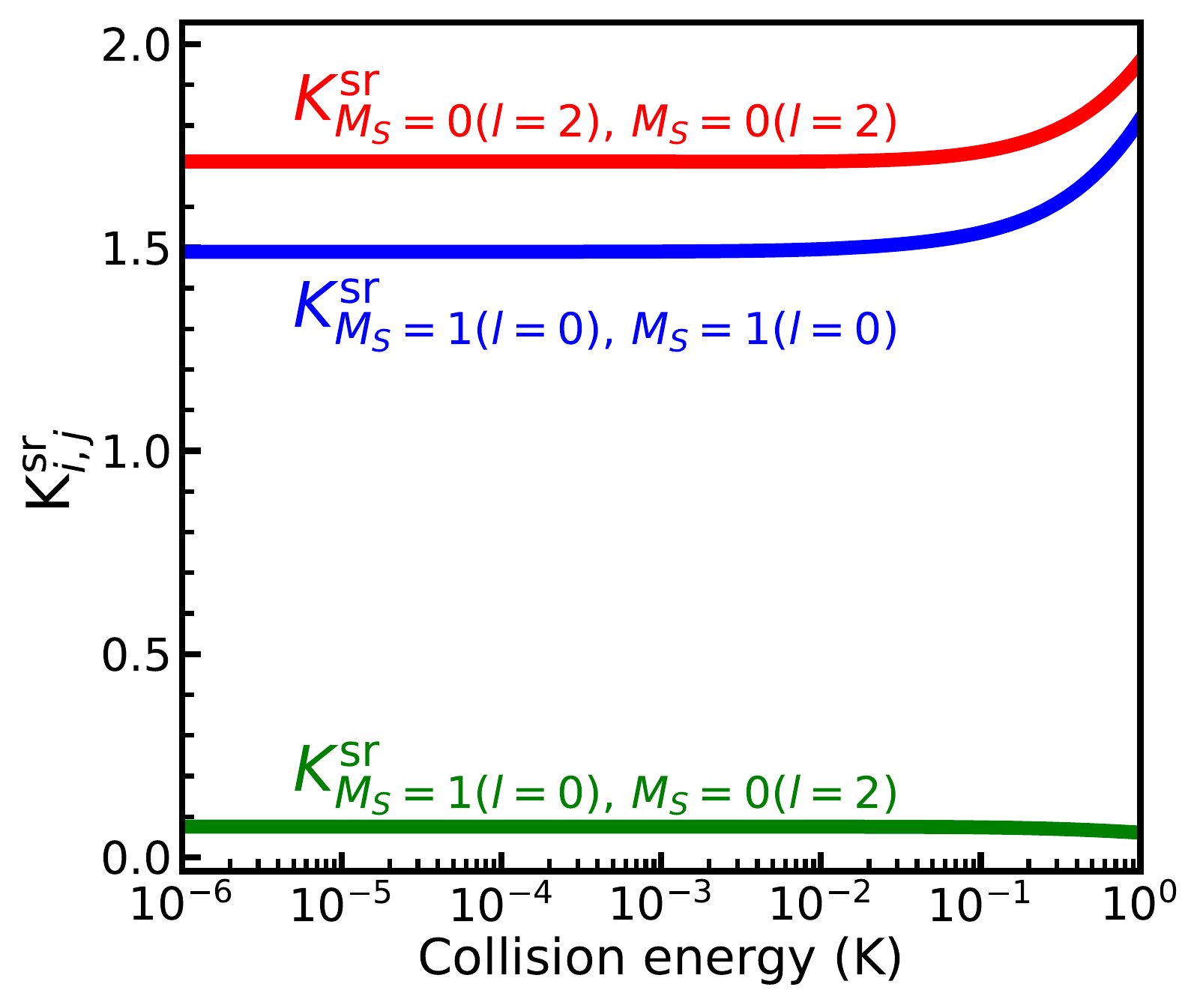}
\end{center}
\caption{Elements of the $2\times2$ short-range $K$-matrix plotted as functions of collision energy at  $B=1000$\,G for the spin relaxation transition $M_S=1$ ($l=0$) $\to$ $M^\prime_S=0$ ($l^\prime=2$) in Mg+$^{14}$NH ($N=0$) collisions. 
}
\label{SM_Fig_3}
\end{figure}
%%%%%%%%%%%%%%%%%%%%%%%%%%%%%%%%%%%%%%%%%%%%%%%%%%%%%%

\begin{center}
\section{\label{sec:SM_Y} Short-range $K$-matrix}
\end{center}

In Fig.~3 of the main text, we show MQDT-FT results for  spin relaxation cross sections in Mg~+~NH collisions obtained using a reduced $2\times2$ short-range $K$-matrix. \Cref{SM_Fig_3} shows the variation of the matrix elements $K_{ij}^\text{sr}$ with collision energy. Both diagonal (blue and red) and off-diagonal (green) elements are constant throughout the low energy region ($E_\text{c} < 10$mK), indicating that 3 parameters are sufficient to describe spin relaxation in cold Mg+NH collisions over a wide range of collision energies spanning 4 orders of magnitude.

We also observed no magnetic field dependence of $K_{ij}^\text{sr}$.
In particular, the off-diagonal matrix element $K_{M_S=1(l=0),M_S=0(l=2)}^\text{sr}$ is conserved to within 6 significant digits in the $B=1-1000$\,G range at $E_\text{c}=10^{-6}$K.
%The stability of the matrix elements of the matrix about energy is displayed in \cref{SM_Fig_3}. Both diagonal (blue and red) and off-diagonal (green) elements behave as constant values through low energy region ($E_\text{c} < 10$mK), indicating the 3 parameters are sufficient to describe the short-range information for this transition in cold and ultracold energy regimes. The stability against the magnetic field is also observed. We observe the conservation of the 6-digits for the off-diagonal matrix element in the $B=1-1000$\,G region with $E_\text{c}=10^{-6}$K. 
\newline

\vspace{0.8cm}
\section{\label{sec:SM_fine} Fine structure transitions}

%%%%%%%%%%%%%%%%%%%%%%%%%%%%%%%%%%%%%%%%%%%%%%%%%%%%%%
\begin{figure}[h!]
\begin{center}
\includegraphics[height=0.27\textheight,keepaspectratio]{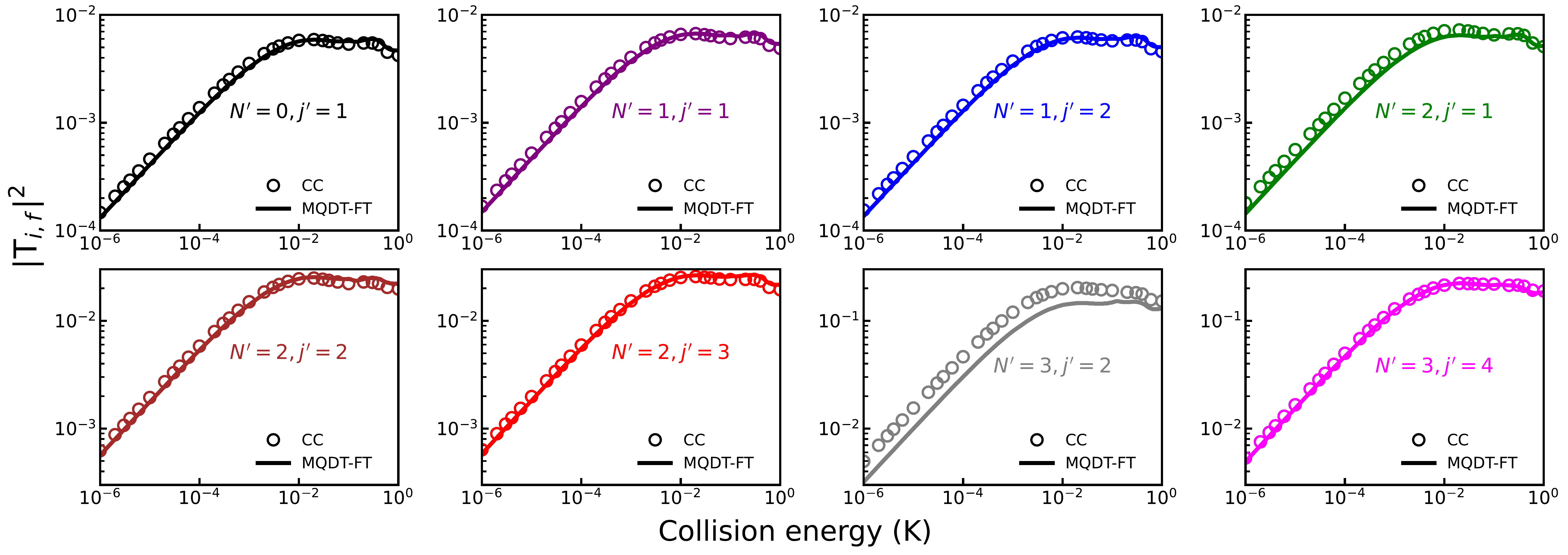}
\end{center}
\caption{
Fine structure transition probabilities in ultracold 
 Mg~+~$^{14}$NH($N=3$, $j=3$) collisions plotted as functions of collision energy for all possible final states and $l=0$. 
}
\label{SM_Fig_4}
\end{figure}
%%%%%%%%%%%%%%%%%%%%%%%%%%%%%%%%%%%%%%%%%%%%%%%%%%%%%%

To calculate the cross sections for fine-structure transitions via the MQDT-FT procedure outlined in the main text, it is necessary to express the short-range $K$-matrix ($\bm{K}^\text{sr}$) in the $|(NS)jlJM\rangle$ basis. However, our short-range $K$-matrix ($\bm{K}_\mathrm{0}^\text{sr}$) is given in the spin-free  $|(Nl)J_rM_r\rangle$ basis. 
These basis sets are related by an analytical recoupling transformation given by Eq.~(2.27) of Ref.~\cite{Corey_83}
\begin{equation}
\begin{split}
K^{\text{sr},J}_{N'Sj'l',NSjl} =  \sum_{J_r=|N-l|}^{N+l} [J_r][j' j]^{1/2} (-1)^{-l'+l+j'-j}   
\begin{Bmatrix} S & N' & j' \\ l' & J & J_r \end{Bmatrix} 
\begin{Bmatrix} S & N  & j  \\ l & J & J_r \end{Bmatrix}
K^{\text{sr},J_r}_{0,N'l',Nl},
\label{eq:7}
\end{split}
\end{equation}
where $\{\}$ denotes a 6-$j$ symbol, $[J_r]=(2J_r+1)$, and  $[j,j']=(2j+1)(2j'+1)$.
This can be interpreted as the frame transformation between the basis sets of $|(NS)jlJM\rangle$ and $|(Nl)J_rM_r\rangle |S,M_S>$.

%In Fig.~4(b) of the main text, we show the cross sections for fine structure  transitions from the $N=3$, $j=3$ ($l=0$) initial state of NH in cold collisions with Mg atoms.
In Fig.~4(a) of the main text, we show the probabilities for fine-structure transitions in cold Mg~+~NH collisions.   %from the $N=3$, $j=3$ ($l=0$) initial state of NH in cold collisions with Mg atoms.
 \Cref{SM_Fig_4} shows an extended sample of the results for all possible final states of NH using the recoupling FT. 
All the transitions accompanied by rotational de-excitation ($N=3$ $\to$ $N^\prime=0,1,$ and $2$) are seen to be accurately described by MQDT-FT. For the $N$-conserving, $j$-changing transitions within the $N=3$ manifold, we observe a slight discrepancy between the CC and MQDT-FT results for $j^\prime=2$. The overall collision energy dependence of this transition is well reproduced by MQDT-FT. 
%On the other hand, we observe excellent agreement for $j^\prime=4$. 
This can be explained by noting that the recoupling transformation (\cref{eq:7}) relies on the electron spin playing a spectator role and neglects the intramolecular spin-spin interaction in NH, which plays an important role in $N$-conserving fine structure transitions. 

%Here, we show the results of the transitions to all possible final states in \cref{SM_Fig_4} using the recoupling technique for the frame transformation of the $K$-matrix. All the transitions accompanied by the rotational relaxations ($N=3$ $\to$ $N^\prime=0,1,$ and $2$) are calculated precisely with the MQDT-FT using the recoupling technique. For the pure $j$-state transitions in $N=3$, we see a slight discrepancy between the CC and MQDT results for $N^\prime=3$, $j^\prime=2$ despite the overall collision energy dependence is reproduced well. On the other hand, we observe excellent agreement for $N^\prime=3$, $j^\prime=4$ as written shown in the main text. It is not surprising if the MQDT-FT does not work for the $N$-conserving transition as long as we use the recouping technique because it relies on the assumption that the spectator role of spin in the rotational (rovibrational) transitions. However, further investigation and analysis may help us to extend the recoupling technique or find an effective method for all the $N$-conserving fine state transitions.  

%\vspace{0.8cm}
\section{\label{sec:SM_FR} MQDT-FT and CC calculations near a Feshbach resonance}

%As an extremely important application of MQDT-FT is the prediction of magnetically tunable Feshbach resonances because many relevant atom-molecule systems for ultracold collisions are expected to exhibit high density of resonances due to the coupling of different rotational states induced by the anisotropy of the interaction potentials. 
%In a forthcoming publication, we will conduct extensive study on the application of MQDT-FT for describing magnetic Feshbach resonances in ultracold atom-molecule collisions. 
%In the present paper (Fig.~4 (b)), we present

Here, we present the details of our preliminary MQDT-FT  calculations of a Feshbach resonance in cold  Mg+NH collisions 
(a more extensive study will be presented in a future publication).
% we will conduct extensive study on the application of MQDT-FT for describing magnetic Feshbach resonances in ultracold atom-molecule collisions.
Because of the low density of magnetic Feshbach resonances in Mg~+~NH collisions and their narrow widths,  locating such resonances is challenging in fully converged CC calculations.  To reduce the computational effort, we used a restricted uncoupled CC basis set with $N_\text{max}=1$ and $l_\text{max}=3$, and scaled the Mg-NH interaction potential by a constant  factor of $\lambda=0.87$.  We note that using the unconverged basis is appropriate for the purpose of comparing MQDT-FT results with full CC calculations \cite{Croft_11}.
The simplified Hamiltonian $\hat{H}_0$ used in the MQDT-FT  calculations is the same as that used in the calculations shown in Figs.~1 and 2 of the main text. 

%While using limited basis sets could affect the position and width of the resonance, we assume, , that is no reason to assume that it would adversely affect the performance of the MQDT-FT approach. \textcolor{magenta}{ I remember you did calculations for other $\lambda$ values, did you also try larger basis sets? If so, it would be nice to mention these test calculations here.}
%we checked that the  reduced basis and potential scaling do
%in these preliminary calculations. 

We observe a single resonance peak at $B\simeq 23.5$~G in both the MQDT-FT and CC calculations (see Fig.~4(b) of the main text).
 The peak disappears on the exclusion of the $N=1$ rotational state from the basis set, confirming that it is indeed a Feshbach resonance due to a closed-channel state in the $N=1$ manifold of NH. Additionally, we found that the observed resonance arises from the quasi-bound state dissociating into the $N=0, \, M_S=1,\, l=2$ channel.

\bibliography{Master.bib,MQDTFT.bib,extra.bib,reply.bib}
\end{document}